%% file: ms.tex
\documentclass{mn2e}
\usepackage[dvipsnames,usenames]{color}
\usepackage{colortbl}
\usepackage{epsf,times,mathptmx,graphicx}
\usepackage{amsmath}
\let\bibhang\relax

\usepackage{natbib}

\newcommand{\etal}{{et al}\/.}
\begin{document}
\title[{\rm Herschel}-ATLAS/GAMA: radio galaxies]{{\it
    Herschel}-ATLAS/GAMA: a difference between star-formation rates in
strong-line and weak-line radio galaxies\thanks{{\it Herschel} is an ESA space observatory with science instruments provided by European-led Principal Investigator consortia and with important participation from NASA.}}
\author[M.J.\ Hardcastle \etal]{M.J.\ Hardcastle$^{1}$,
J.H.Y.\ Ching$^{2}$,
J.S.\ Virdee$^{3}$,
M.J.\ Jarvis$^{1,3,4}$,
S.M.~Croom$^{2}$,
\newauthor
E.M.\ Sadler$^{2}$,
T.\ Mauch$^{1}$,
D.J.B.\ Smith$^{1}$,
J.A.\ Stevens$^{1}$,
M.\ Baes$^{5}$,
I.K.\ Baldry$^{6}$,
\newauthor
S.\ Brough$^{7}$,
A.\ Cooray$^{8}$,
A. Dariush$^{9,10}$,
G.\ De Zotti$^{11,12}$,
S.\ Driver$^{13,14}$,
L.\ Dunne$^{15}$,
\newauthor
S.\ Dye$^{16}$,
S.\ Eales$^{17}$,
R.\ Hopwood$^{10,18}$,
J.\ Liske$^{19}$,
S.\ Maddox$^{15}$,
M.J.\ Micha{\l}owski$^{20}$,
\newauthor
E.E.\ Rigby$^{21}$,
A.S.G.\ Robotham$^{13,14}$,
O.\ Steele$^{22}$,
D.\ Thomas$^{22}$,
and
E.\ Valiante$^{17}$
\\
$^{1}$ School of Physics, Astronomy and Mathematics, University of Hertfordshire, College Lane, Hatfield AL10 9AB\\
$^{2}$ Sydney Institute for Astronomy, School of Physics, University of Sydney, NSW 2006, Australia\\
$^{3}$ Oxford Astrophysics, Denys Wilkinson Building, University of Oxford, Keble Rd, Oxford OX1 3RH\\
$^{4}$ Physics Department, University of the Western Cape, Cape Town, 7535, South Africa\\
$^{5}$ Sterrenkundig Observatorium, Universiteit Gent, Krijgslaan 281 S9, B-9000 Gent, Belgium\\
$^{6}$ Astrophysics Research Institute, Liverpool John Moores University, Twelve Quays House, Egerton Wharf Birkenhead, CH41 1LD\\
$^{7}$ Australian Astronomical Observatory, PO Box 915, North Ryde, NSW 1670, Australia\\
$^{8}$ Department of Physics and Astronomy, University of California, Irvine, CA 92697, USA\\
$^{9}$ Institute of Astronomy, University of Cambridge, Madingley Road, Cambridge, CB3 0HA\\
$^{10}$ Physics Department, Imperial College London, South Kensington Campus, London, SW7 2AZ\\
$^{11}$ INAF-Osservatorio Astronomico di Padova, Vicolo Osservatorio 5, I-35122 Padova, Italy\\
$^{12}$ SISSA, Via Bonomea 265, I-34136 Trieste, Italy\\
$^{13}$ ICRAR M468, The University of Western Australia, 35 Stirling Highway, Crawley, WA 6009, Australia\\
$^{14}$ School of Physics \&\ Astronomy, University of St Andrews, North Haugh, St Andrews, KY16 9SS\\
$^{15}$ Department of Physics and Astronomy, University of Canterbury, Private Bag 4800, Christchurch, New Zealand\\
$^{16}$ School of Physics and Astronomy, University of Nottingham, University Park, Nottingham, NG7 2RD\\
$^{17}$ School of Physics \&\ Astronomy, Cardiff University, The Parade, Cardiff, CF24 3AA\\
$^{18}$ Department of Physical Sciences, The Open University, Milton Keynes MK7 6AA, UK\\
$^{19}$ European Southern Observatory, Karl-Schwarzschild-Str. 2, 85748 Garching, Germany\\
$^{20}$ Scottish Universities Physics Alliance, Institute for Astronomy, University of Edinburgh, Royal Observatory, Edinburgh, EH9 3HJ\\
$^{21}$ Leiden Observatory, PO Box 9513, 2300 RA, Leiden, the Netherlands\\
$^{22}$ Institute of Cosmology and Gravitation, University of Portsmouth, Dennis Sciama Building, Burnaby Road, Portsmouth, PO1 3FX\\
}
\maketitle
\vbox{\vskip -180pt}
\begin{abstract}
We have constructed a sample of radio-loud objects with optical
spectroscopy from the Galaxy and Mass Assembly (GAMA) project over the
{\it Herschel}-ATLAS Phase 1 fields. Classifying the radio sources in
terms of their optical spectra, we find that strong-emission-line
sources (`high-excitation radio galaxies') have, on average, a factor
$\sim 4$ higher 250-$\mu$m {\it Herschel} luminosity than weak-line
(`low-excitation') radio galaxies and are also more luminous than
magnitude-matched radio-quiet galaxies at the same redshift. Using all
five H-ATLAS bands, we show that this difference in luminosity between
the emission-line classes arises mostly from a difference in the
average dust temperature; strong-emission-line sources tend to have
comparable dust masses to, but higher dust temperatures than, radio
galaxies with weak emission lines. We interpret this as showing that
radio galaxies with strong nuclear emission lines are much more likely
to be associated with star formation in their host galaxy, although
there is certainly not a one-to-one relationship between star
formation and strong-line AGN activity. The strong-line sources are
estimated to have star-formation rates at least a factor 3-4 higher
than those in the weak-line objects. Our conclusion is consistent with
earlier work, generally carried out using much smaller samples, and
reinforces the general picture of high-excitation radio galaxies as
being located in lower-mass, less evolved host galaxies than their
low-excitation counterparts.
\end{abstract}
\vskip -60pt
\clearpage
\section{Introduction}
\label{intro}

The relationship between AGN activity and star formation is a complex
one. In order to maintain the observed black hole mass/bulge mass
relationship, black holes must grow as new stars form
\citep[e.g.][]{Magorrian+98} and black hole growth should result in
AGN activity. The generally accepted picture is one in which mergers
trigger both AGN activity and star formation
\citep[e.g.][]{Granato+04,diMatteo+05} and in which the AGN activity,
at some point, shuts down star formation by one of a range of
processes generally referred to as `feedback'
\citep[e.g.][]{Croton+06}. The microphysics of this process presumably
involves the driving of outflows either by luminous quasar activity
\citep[e.g.][]{Maiolino+12} and/or radio jets
\citep[e.g.][]{Hardcastle+12}; an understanding of how feedback operates
in populations of galaxies is crucial to models of galaxy and black
hole evolution.

Radio-loud active galaxies form a particularly interesting
sub-population of AGN in the context of this question. Firstly, they
tend to reside in massive elliptical galaxies, traditionally thought
to be `red and dead' with little or no recent star formation;
secondly, the large amount of kinetic energy that they inject into
their environment means that they must both influence and be
influenced by the galactic environment in which they are embedded.
There is in fact long-standing observational evidence
\citep[e.g.][]{Heckman+86} that some powerful radio galaxies have
peculiar optical morphologies, plausibly the results of mergers with
gas-rich galaxies. In these systems, we might expect AGN activity and
star formation to go hand in hand, although the different timescales
for star formation and AGN triggering will mean that they will not
always be observed together; in radio-quiet systems, there may be
several hundred Myr of delay between the starburst and the peak of AGN
accretion \citep{Wild+10}. In contrast to these objects, we know that
other radio galaxies, often equally powerful when their kinetic powers
can be computed, reside in the centres of rich cluster or group
environments where, on the one hand, gas-rich mergers must be very
rare, and, on the other, the duty cycle of AGN activity must approach
100 per cent to account for the nearly universal detection of radio
sources in these systems \citep[e.g.][]{Eilek+Owen06}. In these
objects, we would be surprised to see evidence for a direct link
between AGN activity and star formation.

It may be possible, as originally suggested by \cite{Heckman+86}, to
understand these apparently contradictory results in the context of a
two-population model of the AGN activity in radio galaxies. The two
populations in question probably correspond quite closely to classes A
and B of \cite{Hine+Longair79}, now known as high-excitation and
low-excitation radio galaxies \citep[hereafter HERGs and LERGs:
  e.g.][]{Laing+94, Jackson+Rawlings97}. In recent years it has become
clear that the differences between these objects are not simply a
matter of emission-line strength but extend to optical
\citep{Chiaberge+02}, X-ray \citep*{Hardcastle+06-2} and mid-IR
\citep*{Ogle+06,Hardcastle+09}. In the vast majority of
LERGs\footnote{The optical emission-line class does not correspond
  completely reliably to other indicators of AGN activity; see
  \cite{Hardcastle+09} for a discussion of some anomalous or
  intermediate objects and \cite{RamosAlmeida+11a} for a particularly
  well-documented `LERG' with a clear heavily absorbed, luminous
  hidden AGN. Emission-line classification clearly does not have a
  one-to-one relationship to radiative efficiency, but, for
  simplicity, in this paper we will continue to refer to LERGs and
  HERGs as though they represent the archetypes of their population.},
there is no evidence for any radiatively efficient AGN activity,
setting aside non-thermal emission associated with the nuclear jet
(e.g. \citealt{Hardcastle+09} and references therein); the AGN power
output is primarily kinetic and we observe it only through the
radiation of the jet and lobes and through the work they do on the
medium in which they are embedded. On the other hand, the HERGs, which
include the traditional classes of narrow-line radio galaxies with
spectra like those of Seyfert 2s, the broad-line radio galaxies and
the radio-loud quasars, behave like textbook AGN with the addition of
jets and lobes. Although LERGs are more prevalent at low radio powers
and HERGs at high powers, both classes are found across the vast
majority of the radio power range and, where they overlap, there is
often no way of distinguishing between the radio structures that they
produce.

The reason for the fundamental differences between the AGN activity in
these two classes of radio source is not clear, but one proposal is
that the differences arise because of different fuelling mechanisms.
In this scenario \citep{Hardcastle+07-2} the LERGs are fuelled
directly from the hot gas halos of their host ellipticals and the
groups and clusters in which they lie, while the HERGs are fuelled,
often at a higher rate, by cold gas, presumably brought into the host
elliptical by mergers or interactions with gas-rich
systems\footnote{It is not yet clear whether the difference in the AGN
  results from the difference in the temperature of the accreted
  material, as proposed by Hardcastle \etal , or simply from the lower
  accretion rates as a fraction of Eddington expected for massive
  black holes being fed at something approximating the Bondi rate in
  the LERGs, as in the models of \cite{Merloni+Heinz08} and as argued
  by \cite{Best+Heckman12}; Mingo \etal\ (in prep) will discuss this
  question in detail. However, the answer to this question makes very
  little difference to the predictions of the model.}. Because the
LERGs dominate the population at low power and low redshift, this
allows a picture in which nearby radio-loud AGN are driven by
accretion of the hot phase and are responsible for balancing its
radiative cooling \citep[e.g.][]{Best+06} while still allowing for
merger- and interaction-driven radio-loud AGN at higher radio
luminosity and/or redshift. This model makes a number of testable
predictions. LERGs will tend to be associated with the most massive
systems, will therefore tend to inhabit rich environments, and will
largely have old stellar populations; as a population, they will
evolve relatively slowly. HERGs can occur in lower-mass galaxies with
lower-mass black holes, provided that there is a supply of (cold)
fuel: we therefore expect them to be in less dense environments, to be
associated with merger and star-formation signatures, to be in less
evolved, lower-mass galaxies and to evolve relatively fast with cosmic
time (since the merger rate was higher in the past). Many of these
predictions have been tested. There is some evidence, particularly at
low redshifts, for a difference in the environments and the masses of
the host galaxies of LERGs and HERGs \citep[][Ching \etal, in prep.]{Hardcastle04,Tasse+08}, and there is strong evidence, also at low
redshifts, for differences in the host galaxy colours in the sense
expected from the model described above
\citep{Smolcic09,Best+Heckman12,Janssen+12}. There is strong evidence
for an increased fraction of signatures of merger or interaction in
the galaxy morphologies of the HERGs with respect to the LERGs
\citep{RamosAlmeida+11b} and with respect to a background galaxy
population \citep{RamosAlmeida+12}. And, most importantly from the
point of view of the present paper, there is direct evidence for
different star-formation histories in the hosts of HERGs and LERGs, in
the sense predicted by the model, i.e. that HERGs show evidence for
more recent star formation both at low redshift
\citep{Baldi+Capetti08} and at $z \sim 0.5$ \citep{Herbert+10}.

Studies of the star formation in the different classes of radio galaxy
have until recently been limited in size because of the techniques and
samples used (e.g. {\it HST} imaging by Baldi \& Capetti, analysis of
optical spectroscopy by Herbert \etal ). Only recently have large
samples begun to be analysed \citep{Best+Heckman12,Janssen+12} and so
far this work has been based only on optical colours at low redshift.
Mid-infrared observations with {\it Spitzer} provide some evidence
that individual HERGs may have strong star formation (e.g. Cygnus A,
\citealt{Privon+12}) but systematic studies of large samples have
generally shown that the luminosity in the mid-IR is dominated by
emission from the AGN itself, by way of the dusty torus \citep[e.g.][]{Hardcastle+09,Dicken+09};
detailed mid-IR spectroscopy in small samples \citep{Dicken+12} has
shown that there is not a one-to-one association between star
formation signatures and AGN activity, but this type of work cannot
easily be extended to very large samples. However, observations of
cool dust in the far infrared (FIR) should, in principle, provide a
very clear way of studying star formation, which should be
uncontaminated by AGN activity, since the emission from the dusty
  torus of the AGN is found to peak in the rest-frame mid-IR
  \citep[e.g.][]{Haas+04}. FIR observations can be carried out simply for
large samples, and the method can extend to relatively high redshifts, with
the only contaminant being emission from diffuse dust heated by the
local interstellar radiation field rather than by young stars
(at least until redshifts become so high that rest-frame mid-IR torus
  emission starts to appear in the observer-frame FIR bands). Earlier
work on far-infrared/sub-mm studies of star formation in samples of radio
galaxies necessarily concentrated on high-redshift objects, in which
emission at long observed wavelengths (e.g. 850 $\mu$m, 1.2 mm)
corresponds to rest-frame wavelengths around the expected peak
  of thermal dust emission \citep[e.g.][]{Archibald+01,Reuland+04} and thus
  applied only to very radio-luminous AGN. Much larger and more
local samples can be studied using the {\it Herschel
  Space Observatory} \citep{Pilbratt+10} and in particular by
wide-field surveys such as the {\it Herschel} Astrophysical Terahertz
Large Area Survey \citep[H-ATLAS;][]{Eales+10}.

In an earlier paper \citep[][hereafter H10]{Hardcastle+10b}, we
studied the FIR properties of radio-loud objects in the
14-square-degree field of the Science Demonstration Phase (SDP)
dataset of H-ATLAS, and showed that, as a sample, their FIR properties
were very similar to those of normal radio-quiet galaxies of similar
magnitude; however, our sample size was small and we were not able to
classify our radio-loud objects spectroscopically. The full `Phase 1'
ATLAS dataset, consisting of three large equatorial regions, gives a
field almost twelve times larger (161 square degrees). Our work on
radio galaxies in the Phase 1 dataset is divided between two papers.
Virdee \etal\ (2012; hereafter V12) use the same sample selection
process as H10, but use the much larger sample available from the
Phase 1 datasets to investigate the relationship between radio
galaxies and normal galaxies in more detail, dividing the radio-loud
sample by properties such as host galaxy mass and radio source size.
In the present paper, we select our sample so as to be able to
classify our radio sources spectroscopically, using data derived from
the Galaxy and Mass Assembly project (\citealt{Driver+09,Driver+11};
hereafter GAMA) and search for differences in the far-infrared and
star-formation properties of HERGs and LERGs.

Throughout the paper we use a concordance cosmology with $H_0 = 70$ km
s$^{-1}$ Mpc$^{-1}$, $\Omega_{\rm m} = 0.3$ and $\Omega_\Lambda =
0.7$. Spectral index $\alpha$ is defined in the sense that $S \propto
\nu^{-\alpha}$.

\section{Sample selection and measurements}

\subsection{The GAMA sample}
The GAMA survey is a study of galaxy evolution using multiwavelength
data and optical spectrosopic data. In phase I of GAMA, target
galaxies are drawn from the SDSSDR6 photometric catalogue in three
individual $12^\circ \times 4^\circ$ rectangles along the equatorial
regions centred at around 9, 12 and 15 hours of right ascension. A
$r$-band magnitude limit of 19.4 was used for the 9 and 15-h fields
while the 12-h field had a deeper 19.8 mag limit \citep{Driver+11}.
The H-ATLAS Phase I data is taken from regions corresponding closely
to these three fields. Reliable spectroscopic redshifts from previous
surveys (e.g. SDSS, 6dF Galaxy Survey, etc.) were used for GAMA
sources that had them. Those without reliable spectroscopic redshifts
from previous surveys were spectroscopically observed on the
Anglo-Australian Telescope (AAT).

We built a sample of candidate radio galaxies by cross-matching the
Faint Images of the Radio Sky at Twenty-cm \citep[FIRST,][]{Becker+95}
catalogue (16 July 2008) with optical sources ($i < 20.5$ mag,
extinction corrected) from the Sloan Digital Sky Survey Data Release 6
\citep[SDSSDR6;][]{SDSSDR6} in all GAMA regions. The full details of
the cross-matching will be described by Ching \etal\ (in prep.), but a
short summary is provided here. The cross-matching firstly involved
grouping FIRST components that were likely subcomponents of a single
optical source (e.g. the core and lobes of a radio galaxy). The
optical counterparts for the groups were matched automatically if they
satisfied certain criteria based on symmetries of the radio sources,
and/or manually when groups were more complex, by overlaying SDSS
images with FIRST and NRAO VLA Sky Survey \citep[NVSS;][]{Condon+98}
contours. Groups that appeared to be separate individual radio sources
were split into appropriate subgroups matched to their individual
optical counterpart. All FIRST components that were not identified as
a possible subcomponent were cross-matched to the nearest SDSS optical
counterpart with a maximum separation of 2.5 arcsec. This process gave
us a sample of 3168 objects with radio/optical identifications. Some
of these objects, predominantly at low redshifts, had spectra from
the SDSS spectroscopic observations; GAMA does not re-observe such
objects. Others were part of the GAMA main sample. To increase the
spectroscopic sample size, we identified galaxies that were not part
of the GAMA main sample (internal data management unit TilingCatv16,
SURVEY\_CLASS$\ne 1$), and observed some of them as spare-fibre
targets during the main GAMA observing programme (see Ching \etal, in
prep., for more details). The resulting sample, by construction,
contained only sources with usable spectra and spectroscopically
determined redshifts ($nQ \ge 3$, from GAMA data management unit
SpecCatv08; \citealt{Driver+11}), and is flux-limited in the radio,
with a lowest 1.4-GHz flux density around 0.5 mJy and most sources
having flux density above 1.0 mJy, as a result of the use of FIRST in
constructing the sample. There were 2559 sources with spectroscopic
redshifts in this parent sample.

\subsection{Spectral classification}

Spectral classification of the objects with radio/optical
identifications was carried out by inspection of their spectra. A
detailed description of the process will be given by Ching \etal\ (in
prep.); here we simply summarize the steps we followed. The emission
line measurements used in this paper for GAMA spectra were made from
the Gas and Absorption Line Fitting (GANDALF; \citealt{Sarzi+06}) code
as part of the GAMA survey (see \citealt{Hopkins+12} for a description
of the GAMA spectroscopy and spectroscopic pipeline), while for SDSS
spectra we used the measurements from the value-added MPA-JHU
emission-line measurements derived from SDSS
DR7\footnote{http://www.mpa-garching.mpg.de/SDSS/DR7/}. Both of these
measurements fit the underlying stellar population before making
emission line measurements, and hence take into account any stellar
absorption. Only high-quality GAMA spectra were used.

We firstly removed Galactic sources by imposing a lower redshift limit
of $z>0.002$. Such objects are classified `Star' and play no further
part in the analysis in this paper. Next, we visually selected objects
with broad emission lines; these are classified `AeB' in this paper,
and are broad-line radio galaxies or radio-loud quasars.

Galaxies that are within $z<0.3$ and have 1.4-GHz luminosity
(hereafter $L_{1.4}$) below $10^{24}$ W Hz$^{-1}$ have a high
probability of having star-formation dominated radio emission (see
e.g. \citealt{Mauch+Sadler07}). For all such objects having
[O{\sc iii}], [N{\sc ii}], H$\alpha$ and H$\beta$ emission lines detected
with a signal-to-noise ratio $>3$, we used a simple line diagnostic
\citep[BPT;][]{Baldwin+81} to classify `pure star-forming galaxies' as
classified by \cite{Kauffmann+03}. These are classed as `SF' in the
following analysis. However, as pointed out by \cite{Best+05}, line
diagnostics alone are not enough to ensure a clean sample of
radio-loud AGN, since the emission lines may arise from a radio-quiet
AGN, while the detected radio emission might arise from star formation
in another region. In addition, the lines required for BPT analysis
are not available at $z>0.3$. We therefore also classified as `SF' any
object whose H$\alpha$ and 1.4-GHz radio emission placed it within
$3\sigma$ of the relation between these two quantities derived by
\cite{Hopkins+03} for star-forming objects. We emphasise that `SF'
objects are not discarded from the analysis at this stage -- therefore
nothing in this classification prejudices the results of the H-ATLAS
analysis.

Finally, we expected the remaining galaxies to be a reasonably robust
sample of radio-loud AGN, possibly contaminated by $z>0.3$ and/or
extremely luminous ($L_{1.4} > 10^{24}$ W Hz$^{-1}$) star-forming objects.
We therefore classified them using a scheme intended to differentiate
between HERG and LERG radio galaxies. Our preliminary classification
was visual, i.e. objects were classed as `Ae' (corresponding to HERGs)
if they showed strong high-excitation lines such as [O{\sc iii}],
[N{\sc ii}], [Mg{\sc ii}], [C{\sc iii}], [C{\sc iv}] or Ly$\alpha$,
and as `Aa' (corresponding to LERG) otherwise, using a similar
classification scheme to that of \cite{Mauch+Sadler07} -- see their
Section 2.5 for more discussion of this approach and its reliability.
However, we then found that the equivalent width of the [O{\sc iii}]
line gave a very similar division between objects with the advantage
of removing the subjective element of the visual classification. In
the final analysis we classified all galaxies with SNR([O{\sc
    iii}])$>3$ and EW([O{\sc iii}])$>5$\AA\ as `Ae' (HERG-like) and
all objects not otherwise classified as `Aa' (LERG-like). The choice
of 5\AA\ as the equivalent-width cut gives the best match to our
preliminary visual analysis, but we verified that small variations in
this choice made little or no difference to the results presented in
the rest of the paper.

A summary of the classification scheme and the number of objects in
the sample in each of the emission-line classes is given in Table
\ref{class}. Throughout the rest of the paper, we retain a distinction
between the {\it observational} classifications (SF, Aa, Ae, AeB) and
the {\it physical} distinction between star-forming non-AGN sources,
LERGs and HERGs; we discuss how well the observational emission-line
classifications map on to the physical distinctions in the course of
the paper, with a summary in Section \ref{discussion}.

\begin{table*}
\caption{The classification scheme used in this paper and the
  number of objects in each class in the H-ATLAS subsample. Also shown
are the numbers of objects after the application of the `SF cut' based
on the radio/FIR relation, as described in the text.}
\label{class}
\begin{tabular}{lllrr}
\hline
Name&Characteristics&RL AGN class&\multicolumn{2}{c}{Number in
  sample}\\
&&&(Total)&(`SF cut' applied)\\
\hline
Aa&AGN spectra with EW([O{\sc iii}])$\le 5$\AA&LERG&1247&1186\\
Ae&AGN spectra with EW([O{\sc iii}])$>5$\AA&HERG/NLRG&199&156\\
AeB&AGN spectra with strong broad high-excitation lines&HERG/BLRG/QSO&187&194\\
SF&Star forming galaxy based on BPT or $H\alpha$-radio correlation&--&191&8\\
\hline
\end{tabular}
\end{table*}

\subsection{{\it Herschel} flux-density measurements}

The classification over the GAMA fields and the removal of stars gives
us 2066 objects, all of which have positions, SDSS identifications,
FIRST flux densities, spectroscopic redshifts from SDSS, GAMA proper or the
spare-fibre programme, and spectroscopic classifications. Our next
step was to extract flux densities for these objects from the H-ATLAS `Phase
1' images. `Phase 1' of H-ATLAS consists of observations of 161 square
degrees of the sky coincident with the GAMA fields, including the much
smaller SDP field discussed by H10; further information on the Phase 1
dataset will be provided by Hoyos et al. and Valiante et al. (in
prep). We discarded all GAMA objects which were outside the area
covered by H-ATLAS (i.e. where flux densities were not available):
this reduced the sample to 1836 objects, and it is this `H-ATLAS
subsample' that we discuss from now on.

H-ATLAS maps the FIR sky with {\it Herschel}'s Spectral and
Photometric Imaging Receiver \citep[SPIRE;][]{Griffin+10} and the
Photodetector Array Camera and Spectrometer
\citep[PACS;][]{Poglitsch+10}. The process of deriving the images used
in this paper is described by \cite{Pascale+11} and \cite{Ibar+10} for
SPIRE and PACS respectively. For each of the objects in our H-ATLAS
subsample we derived the maximum-likelihood estimate of the flux
density at the object position in the three SPIRE bands (250 $\mu$m,
350 $\mu$m and 500 $\mu$m) by measuring the flux density from the
PSF-convolved H-ATLAS images as in H10, together with the error on the
fluxes. We also extracted PACS flux densities and corresponding errors
from the images at 100 and 160 $\mu$m using circular apertures
appropriate for the PACS beam (respectively 15.0 and 22.5 arcsec) and
using the appropriate aperture corrections, which take account of
whether any pixels have been masked. We add an estimated absolute flux
calibration uncertainty of 10 per cent (PACS) and 7 per cent (SPIRE)
in quadrature to the errors measured from the maps for the purposes of
fitting and stacking, as recommended in H-ATLAS documentation, but
this uncertainty is {\it not} included when considering whether
individual sources are detected.

Only 368 of the H-ATLAS subsample (20 per cent) are detected in the
conservative `$5\sigma$' H-ATLAS source catalogue \citep[created as
described by][]{Rigby+11}. This is a similar $5\sigma$
detection fraction to that obtained by H10. We can relax this
criterion for detection slightly, as we know that there are objects
(the host galaxies of the radio sources) at the positions of interest.
A detection criterion of $2\sigma$ implies that 2.3 per cent of
`detected' sources will be spurious, which is acceptable for our
purposes. However, care needs to be taken when applying such a
criterion to the H-ATLAS data. The images at 250, 350 and 500 $\mu$m
are badly affected by source confusion, and this means that the
statistics of the `noise' -- including confusing sources -- are not
Gaussian. We have therefore conservatively determined our $2\sigma$
cutoff by sampling a large number of random background-subtracted
flux densities from the PSF-convolved maps, and determining the flux level
below which 97.7 per cent of the random fluxes lie, to get a flux density
limit which takes account of confusion. This process returns twice the
local r.m.s. noise if the noise is Gaussian, which turns out to be the
case for the PACS data, but gives substantially higher flux density limits of
24.6, 26.5 and 25.6 mJy for the 250, 350 and 500-$\mu$m SPIRE maps
respectively, corresponding to around 3.8 times the local noise
estimates for 250 $\mu$m. These limits are essentially independent of
the local noise estimates (from the noise maps), which is as expected
since the upper tail of the flux density distribution in the maps is dominated
by the effects of confusing sources. In what follows, we say that a
source is `detected' in a given band if it lies above these confusion
limits (for the SPIRE data) or above the standard $2\sigma$ value (for
PACS). By these criteria, 486 sources (26 per cent) are detected at
250 $\mu$m, the most sensitive SPIRE band; the number falls to 244 (13
per cent) at 500 $\mu$m and 328 (18 per cent) at 100 $\mu$m.

We compared this radio-galaxy sample to the sample of V12, which uses
the method described in H10 to select candidate radio galaxies,
requiring a cross-match between the NVSS and the UKIDSS-LAS
\citep{lawrence+07}, over the original 135-square-degree Phase 1
field. 786 of the current sample match objects in the sample of V12,
and for those objects we find good agreement between the NVSS and
FIRST flux densities, suggesting that there is little missing flux. The
objects that are in the H-ATLAS subsample but are not identified as
radio galaxies in the sample of V12 are either not LAS sources or are
faint radio sources that fall below the NVSS flux density limit but are
detectable with FIRST, and so would not be expected to be in our NVSS
catalogue. We conclude that there is good consistency between the
method used here and the method of H10, in the set of objects where
they overlap, and that there is no reason to suppose that the results
are less robust for the population of faint radio sources that we
study for the first time in this paper.

As noted above, our spectroscopically identified sample is not
complete, in the sense that not all objects that would meet the
selection criteria for spectroscopy have high-quality spectra, and
this should be borne in mind in what follows. No selection bias has
been consciously imposed by our choice of objects for spectroscopic
analysis.

\subsection{Luminosity and dust mass calculations}
\label{luminosity}

The rest-frame 1.4-GHz radio luminosity of the sample sources is
calculated from the FIRST 1.4-GHz flux density and the spectroscopic
redshift, assuming $\alpha=0.8$ as in H10. (We comment on constraints
on the spectral index of objects in the sample in the next subsection.)

H10 used integrated FIR luminosities, but these depend very strongly
on the assumptions made about the underlying spectrum, in particular
the $\beta$ and temperature of the modified blackbody model which is
assumed to describe the data. In this paper we instead use the
monochromatic luminosity at rest-frame 250 $\mu$m, $L_{250}$. This has
the advantage that the assumptions we make about the spectrum only
affect the $K$-correction, and so have negligible effect at low redshift. We
still have to make a choice of the spectrum to use for $K$-correction,
since we cannot fit models to the vast majority of our objects. H10
used a modified blackbody with $T = 26$ K, $\beta = 1.5$, but in this
paper we use $T = 20$ K, $\beta = 1.8$, for reasons that will be
justified by temperature fits in Section \ref{temperatures-ind}.

The disadvantage of this approach is that we lose the ability to
estimate the star-formation rate directly from the integrated FIR
luminosity, as we attempted to do in H10: however, the relationships
commonly used to do this \citep[e.g., those given by][]{Kennicutt98}
are calibrated using starburst galaxies and are not necessarily
applicable in the temperature and luminosity range that most radio
galaxies occupy. Instead, we can consider the 250-$\mu$m luminosity as
representing a dust {\it mass} \citep[as in][]{Dunne+11}; the
`isothermal' dust mass, i.e. the mass derived on the assumption of a
single temperature for the dust, is given by
\begin{equation}
M_{\rm iso} = \frac{L_{250}}{4\pi \kappa_{250} B(\nu_{250}, T)}
\label{dustmass}
\end{equation}
where $\kappa_{250}$ is the dust mass absorption coefficient, which
Dunne \etal\ take to be 0.89 m$^2$ kg$^{-1}$, and $B(\nu, T)$ is the
Planck function. It is clear for this mass estimation method, and also
turns out to be the case for the more complex method discussed by
Dunne \etal , that for a roughly constant $T$ we have a linear
relationship between mass and luminosity, while we also expect a
strong correlation between $L_{\rm 250}$ and $T$ for a fixed dust
mass. Moreover, we expect high values of $T$ to be indicators of
strong star formation, independent of $M_{\rm iso}$. In this paper we
will initially use $L_{250}$ to indicate possible differences in star
formation, and use comparisons of fitted temperatures $T$ to confirm
them. Later we will show that $L_{250}$ can be calibrated to give a
quantitative measure of star-formation rate, subject to some important
caveats.

It is important to note that the {\it Herschel} SPIRE PSF has a FWHM
of 18 arcsec at 250 $\mu$m, which corresponds to linear sizes up to
$\sim 150$ kpc at the redshift of the most distant objects in our
sample. As we noted in H10, the luminosities we measure, and any
corresponding dust masses or temperatures, apply not just to the host
galaxy of the radio source but also to its immediate environment. Star
formation associated with a given AGN might actually be taking place
in a merging system or a nearby companion galaxy.

\section{Results}

\subsection{Subsample properties}
\label{blrg}

Table \ref{class} gives the numbers of objects in the H-ATLAS
subsample that fall into the various emission-line classes defined
above. We see that absorption-line only or weak emission-line spectra
(`Aa': unambiguously corresponding to the expected spectra of
`low-excitation' radio galaxies or LERGs) dominate the population.
There are then roughly equal numbers of the `Ae' objects,
corresponding to the high-excitation narrow-line radio galaxies
(HERGs, or NLRGs), broad-line objects (`AeB') and objects classed as
star-forming on the basis of their spectra (`SF').

The redshift distributions within the emission-line classes are
somewhat different. The objects in the Aa and Ae classes have very
similar redshift distributions, with median redshifts around 0.4, as
we might expect for bright galaxies drawn from the parent (SDSS)
sample, and maximum redshift $\sim 1$. We cannot distinguish between
the redshift distributions of the Aa and Ae classes on a
Kolmogorov-Smirnov (KS) test at the $3\sigma$ confidence level. The SF
galaxies have a clearly different distribution, with median $z \sim
0.08$ and maximum $z \sim 0.3$, suggesting that these are mainly
local, fainter galaxies \citep[as expected from the known different
luminosity functions of the AGN and SF populations; see][]{Mauch+Sadler07}. We retain the SF objects in the sample so as not to
exclude the possibility, at this stage, that some are NLRG with strong
star formation. The broad-line objects have a much wider redshift
distribution, with median $z \sim 1.3$ and maximum $z \sim 3.7$. These
objects are clearly mostly quasars that are in the sample due to their
bright AGN emission. Similarly, if we consider the radio flux density
distributions, we cannot distinguish between the Aa or Ae classes at
high confidence with a KS test, but the SF objects have a significantly
different flux distribution from the Aa and Ae, tending to have
fainter radio flux densities.

The AeB objects are systematically very much brighter in the radio,
suggesting that the combination of radio and optical selection for
these quasars is picking up strongly beamed objects, and this is true
even if we consider only the $z<1$ subsample of the AeB objects. Among
other things, this means that we need to be alert to the possibility
of non-thermal contamination in the {\it Herschel} bands. To check
this, we cross-matched the objects in our sample to the GMRT catalogue
of Mauch \etal\ (in prep.), who have imaged the majority of the Phase
1 area at 325 MHz, using a simple positional matching algorithm with a
maximum offset of 5 arcsec. A total of 536/1836 objects have
counterparts in the GMRT catalogue; the low matching fraction reflects
the incomplete sky coverage and variable sensitivity of the GMRT
survey, as described in detail by Mauch et al. Nevertheless, we can
look for spectral index differences in the matching objects. The
number of cross-matches, together with the mean spectral index and the
values at the 10th and 90th percentile, are tabulated as
a function of emission-line class in Table \ref{spix}.

\begin{table*}
\caption{Spectral indices between 325 MHz and 1.4 GHz for sample
  sources with GMRT survey counterparts}
\label{spix}
\begin{center}
\begin{tabular}{lrrrr}
\hline
Source type&Number of matches&Mean spectral index&10th percentile&90th
percentile\\
\hline
All&536&0.70&0.27&1.09\\
SF&45&0.88&0.58&1.34\\
Aa&356&0.77&0.37&1.09\\
Ae&55&0.68&0.34&0.99\\
AeB&80&0.50&$-0.03$&1.12\\
\hline
\end{tabular}
\end{center}
\end{table*}

\begin{figure}
\epsfxsize 8.5cm
\epsfbox{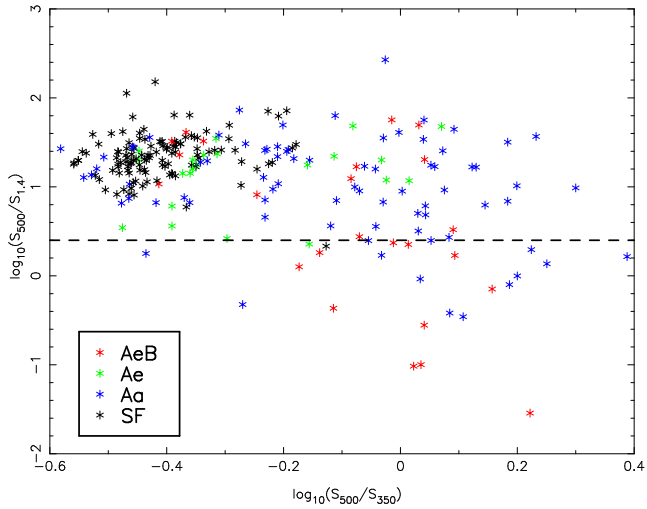}
\caption{Radio sources with a 500-$\mu$m detection plotted on the
  diagnostic plot of \cite{Lopez-Caniego+12}. Colours indicate
  different emission-line classes. The dashed line indicates the
  threshold in FIR/radio radio used by \cite{Lopez-Caniego+12}; below
  this line, synchrotron emission might be bright enough to affect the
  SPIRE bands.}
\label{lopezplot}
\end{figure}

While the large number of non-detections in the GMRT survey means that
we cannot carry out a detailed analysis, we note first of all that the
mean spectral index of detected sources is close enough to our
previously adopted value of 0.8 that our $K$-correction in the radio
will not be badly in error, and secondly that the mean spectral index
of the AeB objects is very much flatter than any of the other
emission-line classes, although there is still clearly a population of
steep-spectrum AeBs. Given that the detected objects are likely to be
biased, if at all, towards the steep-spectrum end of the intrinsic
distribution, it seems likely that the AeB objects contain a
significant number of flat-spectrum quasars.

We investigated this issue further by considering the diagnostic
methods used by \cite{Lopez-Caniego+12} in searching for blazars. They
relied on detections at 500 $\mu$m, and, as noted above, only a small
fraction of our sources have $2\sigma$ detections at that band. We
plotted the sources that do on the diagnostic radio/FIR colour-colour
diagram used by \cite{Lopez-Caniego+12}, which is intended to search
for non-thermal contamination in the SPIRE bands; the result is shown
in Fig.\ \ref{lopezplot}. We see that of the 24 AeB objects with
500-$\mu$m detections, about half lie in the region occupied by the
L\'opez-Caniego blazar candidates in which synchrotron emission might
affect the SPIRE bands, a much higher fraction than for any
other emission-line class. While red 500/350-$\mu$m colours may just
be an indication of low dust temperatures, and the AeB sources have higher
redshifts than the comparison objects, this is a further sign that the
AeBs cannot safely be merged with the Ae objects in what follows.

\subsection{{\it Herschel} and radio luminosity}

\begin{figure*}
\epsfxsize 14cm
\epsfbox{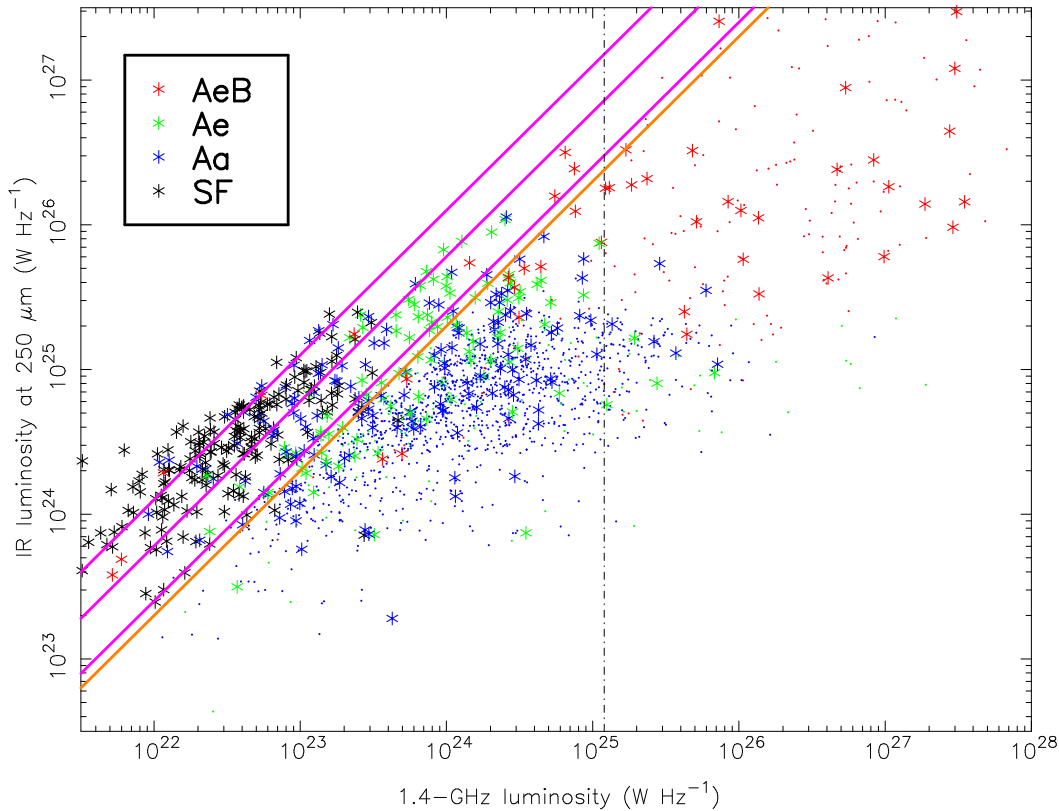}
\caption{250-$\mu$m luminosity against radio luminosity for
  all the objects in the sample. Stars indicate {\it Herschel}
  detections at $2\sigma$ or better as defined in the text, points
  show $2\sigma$ upper limits in IR luminosity derived from the
  confusion limit. Colours correspond to emission-line classes as
  follows: Aa, blue; Ae, green; AeB, red; SF, black. The solid magenta
  lines indicate the expected radio-FIR correlation for star-forming
  objects, $q_{\rm 250} = 1.78$, and the approximate scatter about
  this relation, $1.4 < q_{250} < 2.1$ \citep[from][]{Jarvis+10}.
  The solid orange line shows our adopted `SF cut' at $q_{250}=1.3$, and the
    dot-dashed vertical line shows the nominal FRI/FRII break luminosity.}
\label{llplot}
\end{figure*}

Fig.\ \ref{llplot} shows the IR luminosity, $L_{250}$, against the
radio luminosity for all the objects in the sample. This plot shows
several important features of the sample. First, we note that the vast
majority of the broad-line objects (in red) lie at the very
high-luminosity end of the plot, presumably due to their high
redshifts. As we noted above that some of these objects may well have
FIR fluxes contaminated by non-thermal emission, and as their high
redshift makes it difficult to compare them with radio galaxies in any
case, we exclude them from further analysis.

\begin{figure*}
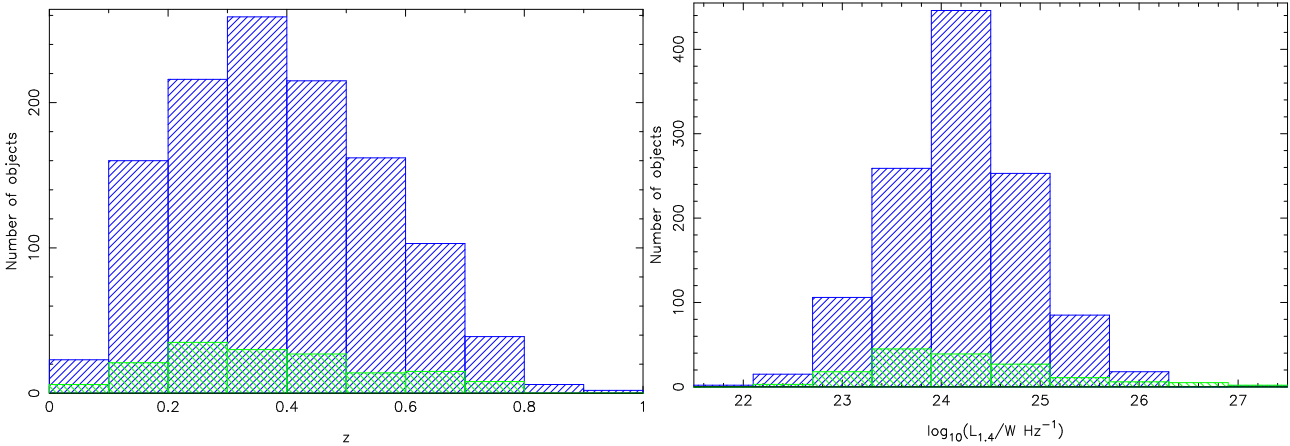

\epsfxsize 8.5cm
\epsfbox{zdist.eps}
\epsfxsize 8.5cm
\epsfbox{ldist.eps}
\caption{The distribution of (left) redshift and (right) radio
  luminosity in the Aa (blue) and Ae (green) objects after the SF cut.
  The redshift and luminosity distributions of the samples are very
  similar.}
\label{lzdist}
\end{figure*}

Fig. \ref{llplot} also shows the expected linear radio-FIR correlation
for star-forming objects (magenta lines), together with the dispersion
seen in that relationship, based on the parameter $q_{250}$, which is
defined as $\log_{10}(L_{250}/L_{1.4})$ \citep{Jarvis+10}. We see that
objects classed as SF on the basis of their emission-line properties
or their radio/H$\alpha$ relation (black points) almost all lie in
this region of the plot and close to the best-fitting line; there is
some positive deviation above the line at low luminosities/redshifts,
but this was also seen by Jarvis \etal . However, at higher
luminosities, a number of objects of other emission-line classes also
fall in the star-forming region, meaning that their radio emission is
not bright enough to definitively classify them as radio galaxies.
Conservatively, every object that lies in the star-forming region of
this plot should be excluded from a discussion of the FIR properties
of radio galaxies; following V12, we adopt a cut at $q_{250}\ge 1.3$.
The numbers of sources remaining, if these objects are excluded, are
given in Table \ref{class}. The vast majority of the SF objects are
removed by the cut (hereafter the `SF cut'), and although some of the 8 remaining sources may
be radio galaxies which have erroneously been classified as SF, we
conservatively exclude them from subsequent analysis (given the small
numbers involved, including them as though they were Ae objects would
have little effect on our results).

Considering only the remaining objects, which we expect to be
radio-loud AGN, we see that these span a very large range in radio
luminosity, from $10^{22}$ to $10^{27}$ W Hz$^{-1}$ if we ignore the
broad-line objects. The vast majority of these lie below the nominal
FRI-FRII luminosity divide \citep{Fanaroff+Riley74} of $1.2\times
10^{25}$ W Hz$^{-1}$ (plotted on Fig.\ \ref{llplot} for reference) and
so would normally be classed as low-luminosity radio galaxies, though
we emphasise that the FRI/FRII division is a morphological one and we
have made no attempt to classify these objects morphologically. In
terms of our observational emission-line classifications, we see that
Aa objects dominate numerically by a large factor, but that there are
Ae objects at all powers. Assuming that Aas trace LERGs and Aes HERGs,
this is consistent both with what is seen in brighter radio-selected
samples at low redshift \citep[see, e.g.,][]{Hardcastle+09} and with
the work of \cite{Best+Heckman12} over a comparable luminosity range.
The Aa and Ae objects left after the SF cut has been made have
redshift and radio luminosity distributions that are indistinguishable
on a KS test (Fig.\ \ref{lzdist}), but this is not surprising, since
the differences in the slope of the luminosity function for the two
populations, leading to the dominance of HERGs at high luminosities,
start to become significant only at $L_{1.4} > 10^{25}$ W Hz$^{-1}$
\citep{Best+Heckman12}, where we have relatively few sources.

\subsection{LERG/HERG comparisons and stacking}
\label{stacking}

Some differences between the FIR properties of the Aa and Ae objects
after the SF cut are immediately obvious on inspection of the data.
For example, 53/156 (34 per cent) of the Ae objects are detected at
the $2\sigma$ level or better at 250 $\mu$m, while only 93/1186 (8
per cent) of the Aa objects are detected at this level.

\begin{figure*}
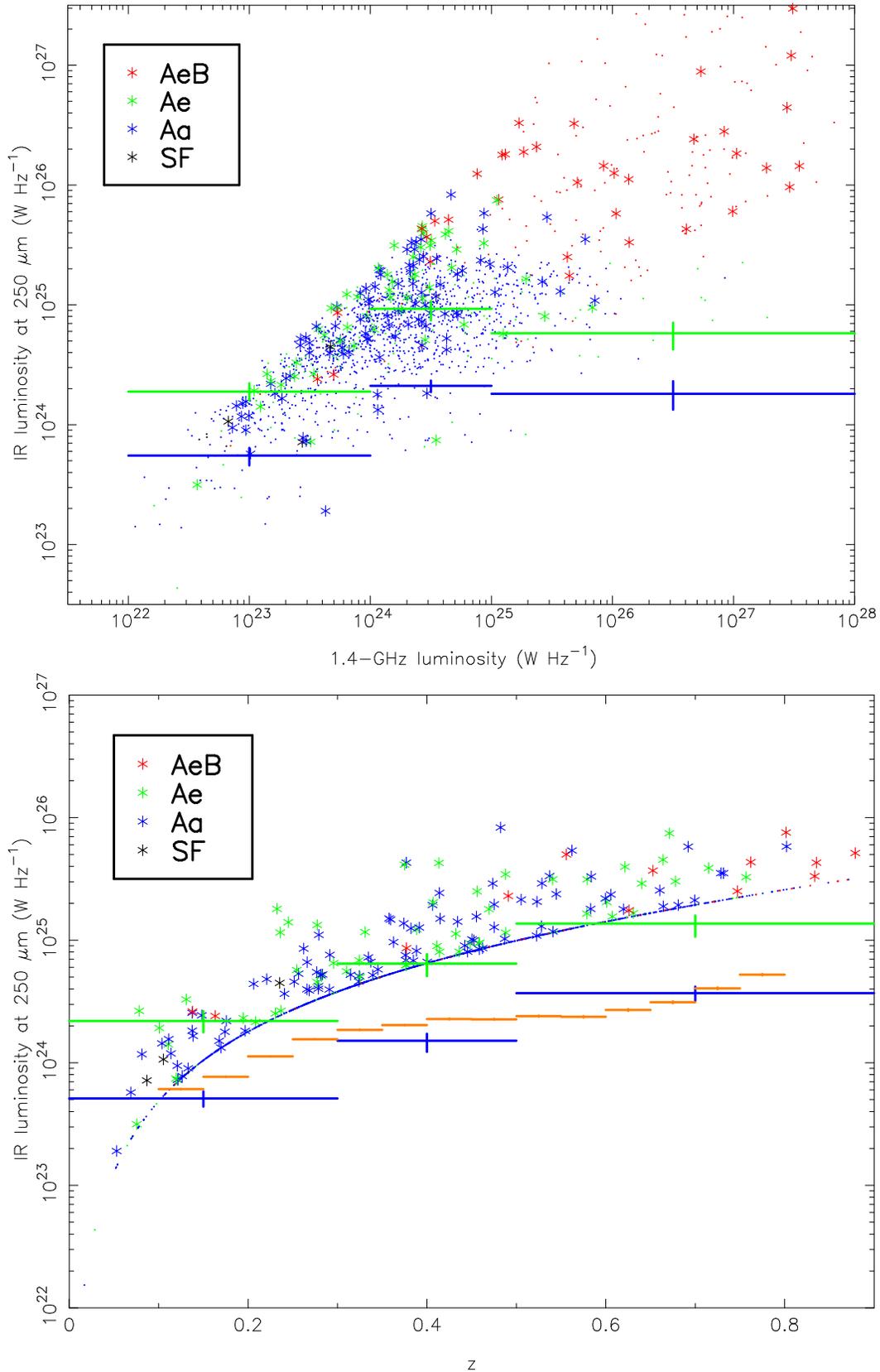

\epsfxsize 14cm
\epsfbox{ll-j-bins2.eps}
\epsfxsize 14cm
\epsfbox{lz-j-bins.eps}
\caption{Comparisons between the IR luminosities of the Aa and Ae
  objects, after excluding objects near the star-formation line, as a
  function of (top) radio luminosity and (bottom) redshift.
  Symbols and colours as for Fig.\ \ref{llplot}. The
  large crosses indicate the results of stacking the IR luminosities
  of all Ae (green) and Aa (blue) objects in the corresponding radio
  luminosity or redshift range. Other types of object are plotted but
  not stacked. The orange bins indicate stacking of comparison
  galaxies, as described in the text.}
\label{compar}
\end{figure*}

Overall, both individual sub-samples still being considered (i.e. Aa
and Ae after SF cut) are significantly detected with respect to the
background at all three {\it Herschel}-SPIRE bands. We follow H10 in
testing this with a KS test on the distribution of flux densities compared to
random flux densities from the field; the highest null hypothesis probability
is $6.6 \times 10^{-8}$ for the Aa sources at 500 $\mu$m,
corresponding to a {\it minimum} significance of $5.5\sigma$ for both
classes and all SPIRE bands, and the significance is much higher at
250 $\mu$m. For the two PACS bands, the Aa sub-sample
is detected at around 98 per cent confidence (i.e. a marginal
detection, $2.3\sigma$) but the Ae objects are significantly detected
with a null hypothesis probability around $2\times 10^{-7}$
($5.2\sigma$).

We are therefore able to adopt the approach of H10 and divide our
sources into luminosity and redshift bins for a stacking analysis.
Since we have relatively few Ae sources, we use only three bins in
both, ensuring that the highest-luminosity bin includes all sources
above the nominal FRI/II luminosity boundary at
$L_{1.4} = 1.2\times 10^{25}$ W Hz$^{-1}$, which, as noted above,
can roughly be taken to separate `low-power' and `high-lower' radio
galaxies. We then used KS tests to see whether these subsamples were
detected (distinguished from the background flux density distribution) at each
of the five H-ATLAS wavelengths. The results are shown in Tables
\ref{zstest} and \ref{lstest}.

\begin{table*}
\caption{Mean bin flux densities and K-S probabilities that the {\it
    Herschel} fluxes of objects in redshift bins (after the SF cut) are
  drawn from the background distribution, as a function of
  emission-line class and wavelength. Low probabilities (below 1 per
  cent) imply significant differences between the bin being considered
  and the distribution of flux densities measured from randomly
  selected positions in the sky, as described in the text. Note that
  the bins do not include quite all the objects in the sample.}
\label{zstest}
\begin{tabular}{llrrrrrrrrrrr}
\hline
Class&$z$ range&Objects&\multicolumn{5}{c}{Mean bin flux density
  (mJy)}&\multicolumn{5}{c}{K-S probability (\%)}\\
&&in bin&\multicolumn{3}{c}{SPIRE bands}&\multicolumn{2}{c}{PACS bands}&\multicolumn{3}{c}{SPIRE bands}&\multicolumn{2}{c}{PACS bands}\\
&&&250 $\mu$m&350 $\mu$m&500 $\mu$m&100 $\mu$m&160 $\mu$m&250 $\mu$m&350 $\mu$m&500 $\mu$m&100 $\mu$m&160 $\mu$m\\
\hline
Aa  & 0.00 -- 0.30 & 399  & $5.5 \pm 0.3$  & $1.0 \pm 0.4$  & $2.5 \pm 0.4$  & $2.7 \pm 1.5$  & $8.7 \pm 1.9$ & $<10^{-3}$ & $<10^{-3}$ & 0.3 & 2.2 & 3.6 \\
 & 0.30 -- 0.50 & 475  & $4.7 \pm 0.3$  & $0.7 \pm 0.3$  & $2.1 \pm 0.4$  & $3.1 \pm 1.4$  & $7.0 \pm 1.7$ & $<10^{-3}$ & 0.004 & 0.09 & 46.1 & 9.9 \\
 & 0.50 -- 0.90 & 310  & $5.3 \pm 0.4$  & $1.4 \pm 0.4$  & $2.6 \pm 0.5$  & $0.2 \pm 1.7$  & $5.9 \pm 2.1$ & $<10^{-3}$ & $<10^{-3}$ & 0.01 & 41.1 & 35.4 \\
Ae  & 0.00 -- 0.30 &  62  & $27.9 \pm 0.9$  & $10.2 \pm 0.9$  & $6.4 \pm 1.1$  & $68.2 \pm 3.8$  & $50.3 \pm 4.6$ & $<10^{-3}$ & $<10^{-3}$ & 0.05 & $<10^{-3}$ & 0.06 \\
 & 0.30 -- 0.50 &  57  & $22.8 \pm 0.8$  & $8.1 \pm 0.9$  & $3.8 \pm 1.1$  & $25.6 \pm 3.8$  & $30.7 \pm 4.7$ & $<10^{-3}$ & 0.005 & 2.1 & 0.3 & 1.5 \\
 & 0.50 -- 0.90 &  37  & $20.4 \pm 1.0$  & $9.8 \pm 1.2$  & $7.6 \pm 1.4$  & $23.5 \pm 5.0$  & $14.2 \pm 6.1$ & $<10^{-3}$ & 0.002 & 0.10 & 11.9 & 20.8 \\
\hline
\end{tabular}
\end{table*}

\begin{table*}
\caption{Mean bin flux densities and K-S probabilities that the {\it
    Herschel} fluxes of objects in luminosity bins (after the SF cut)
  are drawn from the background distribution, as a function of
  wavelength. Notes as for Table \ref{zstest}.}
\label{lstest}
\label{ks-l}
\begin{tabular}{llrrrrrrrrrrr}
\hline
Class&Range in&Objects&\multicolumn{5}{c}{Mean bin flux density (mJy)}&\multicolumn{5}{c}{K-S probability (\%)}\\
&$L_{1.4}$&in bin&\multicolumn{3}{c}{SPIRE bands}&\multicolumn{2}{c}{PACS bands}&\multicolumn{3}{c}{SPIRE bands}&\multicolumn{2}{c}{PACS bands}\\
&&&250 $\mu$m&350 $\mu$m&500 $\mu$m&100 $\mu$m&160 $\mu$m&250 $\mu$m&350 $\mu$m&500 $\mu$m&100 $\mu$m&160 $\mu$m\\
\hline
Aa  & 22.0 -- 24.0  & 457  & $4.7 \pm 0.3$  & $0.5 \pm 0.3$  & $1.9 \pm 0.4$  & $3.2 \pm 1.4$  & $6.9 \pm 1.8$ & $<10^{-3}$ & 0.002 & 0.2 & 2.1 & 14.5 \\
 & 24.0 -- 25.0  & 589  & $5.7 \pm 0.3$  & $1.5 \pm 0.3$  & $2.6 \pm 0.3$  & $2.4 \pm 1.2$  & $7.5 \pm 1.5$ & $<10^{-3}$ & $<10^{-3}$ & $<10^{-3}$ & 49.6 & 8.1 \\
 & 25.0 -- 28.0  & 140  & $3.6 \pm 0.5$  & $0.7 \pm 0.6$  & $2.8 \pm 0.7$  & $-1.7 \pm 2.5$  & $7.4 \pm 3.1$ & 0.006 & 0.4 & 1.8 & 34.0 & 19.8 \\
Ae  & 22.0 -- 24.0  &  71  & $19.0 \pm 0.8$  & $6.3 \pm 0.9$  & $3.7 \pm 1.0$  & $27.2 \pm 3.4$  & $30.7 \pm 4.2$ & $<10^{-3}$ & $<10^{-3}$ & 1.7 & 0.006 & 2.0 \\
 & 24.0 -- 25.0  &  57  & $35.3 \pm 0.8$  & $14.2 \pm 0.9$  & $8.0 \pm 1.1$  & $69.2 \pm 4.0$  & $52.5 \pm 4.8$ & $<10^{-3}$ & $<10^{-3}$ & 0.01 & 0.002 & 0.09 \\
 & 25.0 -- 28.0  &  28  & $15.2 \pm 1.2$  & $7.2 \pm 1.3$  & $6.4 \pm 1.6$  & $24.3 \pm 5.6$  & $7.8 \pm 7.0$ & $<10^{-3}$ & 0.004 & 0.4 & 4.0 & 94.5 \\
\hline
\end{tabular}
\end{table*}

As found by H10, the detection of all our subsamples is best at 250
$\mu$m, although with this larger sample most bins are significantly
detected at 350 $\mu$m as well (see Tables \ref{zstest} and
\ref{lstest}). 500-$\mu$m detections are less robust, and only the
low-redshift Ae subsamples are significantly detected in the PACS
bands. We therefore rely on the 250 $\mu$m flux densities for our first
estimate of luminosities. With the division of the samples into the
two emission-line classes, we can see that the mean 250-$\mu$m flux
density for the Ae objects is much higher than for the Aa objects in
every bin.

Our stacking analysis follows the method of H10; we determine the
luminosity for each source from the background-subtracted flux
density, even if negative, on the grounds that this is the
maximum-likelihood estimator of the true luminosity, and take the
weighted mean within a bin to estimate stacked bin luminosities.
Unlike H10, we determine errors on the bins by bootstrapping, having
verified that this method gives very similar results to the much more
time-consuming and complex method used in the earlier paper. As
Fig.\ \ref{compar} shows, we find a clear and significant difference
between the FIR luminosities of the Ae and Aa objects in every bin in
either radio luminosity and redshift.

\subsection{Comparison with normal galaxies}
\label{normal}

\begin{figure*}
\epsfxsize 17cm
\epsfbox{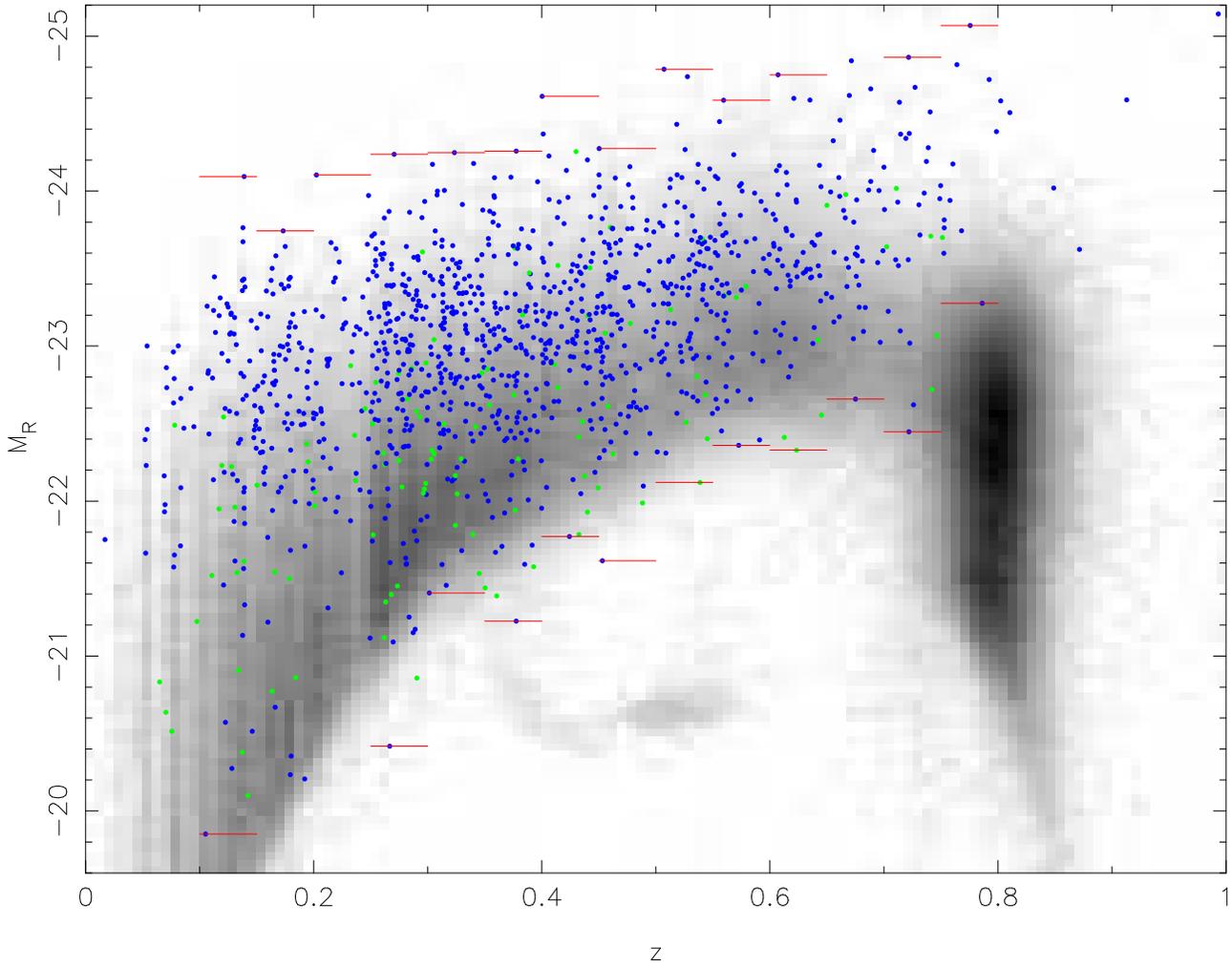}
\caption{The $M_R$-redshift distribution ($K$-corrected as described
  in the text) of sample objects in the Aa (blue) and Ae (green)
  classes, after the `SF cut', plotted on top of the same information
  for the comparison galaxies discussed in the text (greyscale shows a
  density plot with square-root transfer function). Red lines show the
  range of absolute magnitudes of the radio-selected sample, used to
  generate the comparison galaxy sample discussed in the text. The
  peak in redshift around $z\approx 0.8$ is probably an artefact of
  the photometric redshifting, but does not affect our comparison
  sample.}
\label{rz}
\end{figure*}

\begin{figure*}
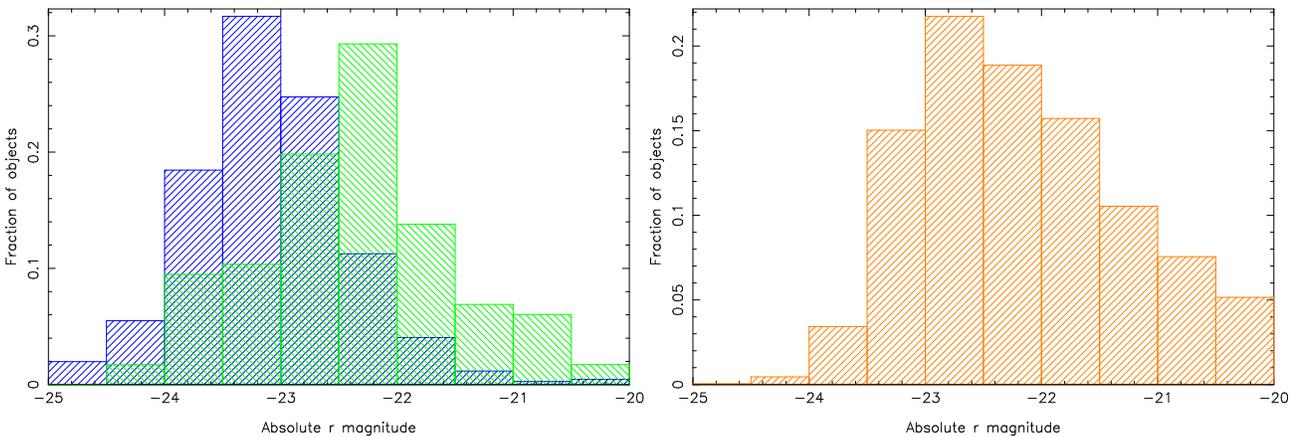

\epsfxsize 8.5cm
\epsfbox{amhist.eps}
\epsfxsize 8.5cm
\epsfbox{amhist-cg.eps}
\caption{Histogram of (left) the absolute $r$ magnitudes of Aa (blue) and Ae (green)
  galaxies in the sample, after application of the SF cut, and (right)
  the absolute magnitudes of the comparison objects after the
  magnitude range selection. Aa objects
  are generally more massive galaxies than Aes, but the distributions have
  substantial overlap; the distribution of comparison galaxy absolute
  magnitudes has an intermediate peak.}
\label{amhist}
\end{figure*}

As a result of our procedure for generating our radio galaxy catalogue
in H10, we automatically had a comparison galaxy population. There is
no equivalent in the present work, in the sense that there is no
galaxy population selected in the same way as the radio-loud objects.
Spectroscopic redshifts for the GAMA sample run out at around $z \sim
0.6$ because of the magnitude limit used by GAMA \citep{Driver+11}
while our spectroscopic sample extends to fainter galaxies and higher redshifts.
However, a rough comparison with radio-quiet galaxies is useful to put
our results in context. We therefore constructed a comparison galaxy
sample as follows:
\begin{enumerate}
\item We based the sample on the galaxy catalogue over the Phase 1
  fields provided as part of the H-ATLAS data release, constructed in
  the manner described by \cite{SmithD+11}, and selected galaxies that
  had either a spectroscopic redshift (from GAMA or SDSS) or a
  photometric redshift with nominal error $<0.1$, had measured SDSS
  $r$ and $i$ magnitudes, and were not point-like in $r$ or $i$.
\item From this sample we took all objects which lay on the observed H-ATLAS
  fields and measured their background-subtracted 250-$\mu$m flux densities as
  for the radio galaxies (giving 318,244 objects in total, the vast
  majority with only photometric redshifts). We excluded at this point
  all objects that formed part of the radio-galaxy sample.
\item We $K$-corrected the $r$-band absolute magnitudes of the radio
  galaxies and the comparison sample to $z=0$ using {\sc kcorrect} v.\ 4.2
  \citep{Blanton+07}.
\item Comparing the range of $r$-band absolute magnitude in the
  radio-galaxy sample with that in the comparison galaxies
  (Fig.\ \ref{rz}) we saw that the radio-selected objects tend to be
  bright galaxies at all redshifts. At lower $z$ there is a tendency
  for the Ae galaxies to be fainter than the Aa (as seen by, for
  example, \citealt{Tasse+08} and \citealt{Best+Heckman12}; note that our
  sample is not complete, which reduces the extent to which we can
  draw conclusions from this observation), but they occupy similarly
  bright galaxies at high $z$ (Fig.\ \ref{amhist}).
\item Clearly for even a
  rough comparison we should compare the radio galaxies with optical
  objects of comparable magnitudes. In each of 14 bins of width
  $\Delta z =0.05$ between $z=0.1$ and $z=0.8$, we selected only the
  comparison galaxies that lay in the absolute $r$ magnitude range
  spanned by the Aa and Ae objects (post-SF cut) in that redshift
  range.
\end{enumerate}

We then stacked the {\it Herschel} FIR luminosities of the galaxies in
those 14 bins, deriving them from the 250-$\mu$m flux densities on the
assumption $T = 20$ K, $\beta = 1.8$ in the same way as for the radio
galaxies. These stacks are plotted as a function of $z$ on
Fig.\ \ref{compar}. We emphasise that this is intentionally a crude
comparison: for example, we could also have performed a colour
selection on the comparison galaxies, but this would have involved an
investigation of the optical colours of the Aa and Ae objects,
accounting for possible AGN contamination, which we wish to defer to a
later paper (Ching \etal, in prep.); similarly, we are not attempting
to separate spirals and ellipticals in the comparison sample, and we
have made no attempt to match the actual distributions of magnitudes
(and thus stellar masses) of the comparison galaxies within the
absolute magnitude ranges used (Fig. \ref{amhist}). Our optical
selection, which is required to allow matching to the radio galaxies,
also potentially biases us against the most strongly star-forming
radio-quiet galaxies, which will tend to be more dust-obscured.
However, the result of the comparison is clear. The luminosities for
the comparison galaxies tend to lie in between those for the Aa and
the Ae radio-loud objects; thus, in the redshift range where we have
data, Ae galaxies are on average {\it more} luminous in the FIR than
the average galaxy of comparable optical magnitude at a given
redshift, and Aa galaxies are on average {\it less} luminous, at
  least in the low-redshift bins where a comparison is possible. We note
that the comparison sample, though differently selected, is behaving
in a very similar manner to that of H10 in terms of its FIR luminosity
as a function of redshift; but now that we have a larger sample and
can separate radio galaxies by emission-line class, we are able to see
differences between the radio-loud and radio-quiet populations.

\subsection{Individual dust temperatures and masses}
\label{temperatures-ind}

Our large sample and the availability of the PACS data allow us to
investigate the temperatures and isothermal dust masses (Section
\ref{luminosity}) of radio-selected objects for the first time.

\begin{figure*}
\epsfxsize 8.5cm
\epsfbox{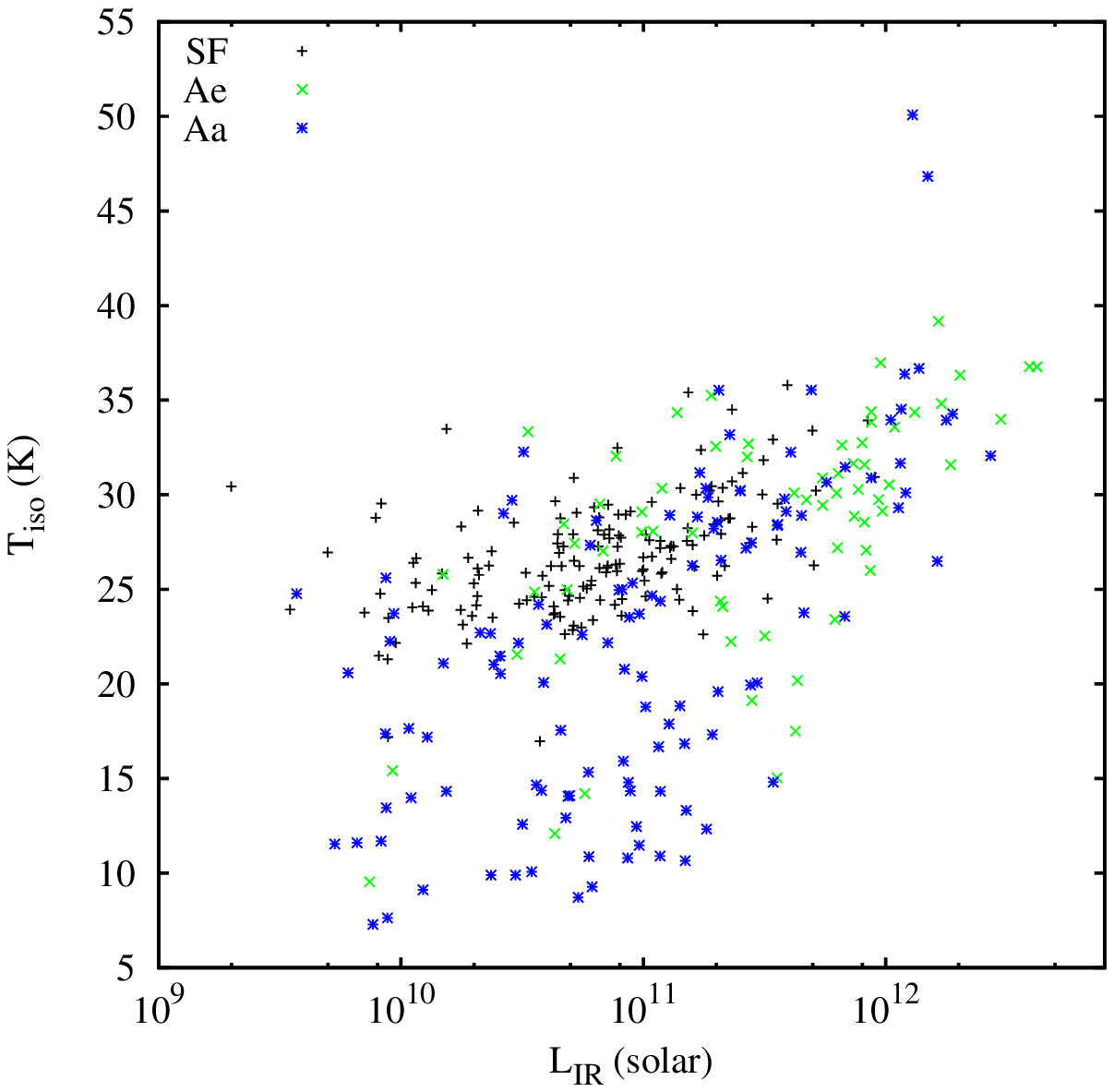}
\epsfxsize 8.5cm
\epsfbox{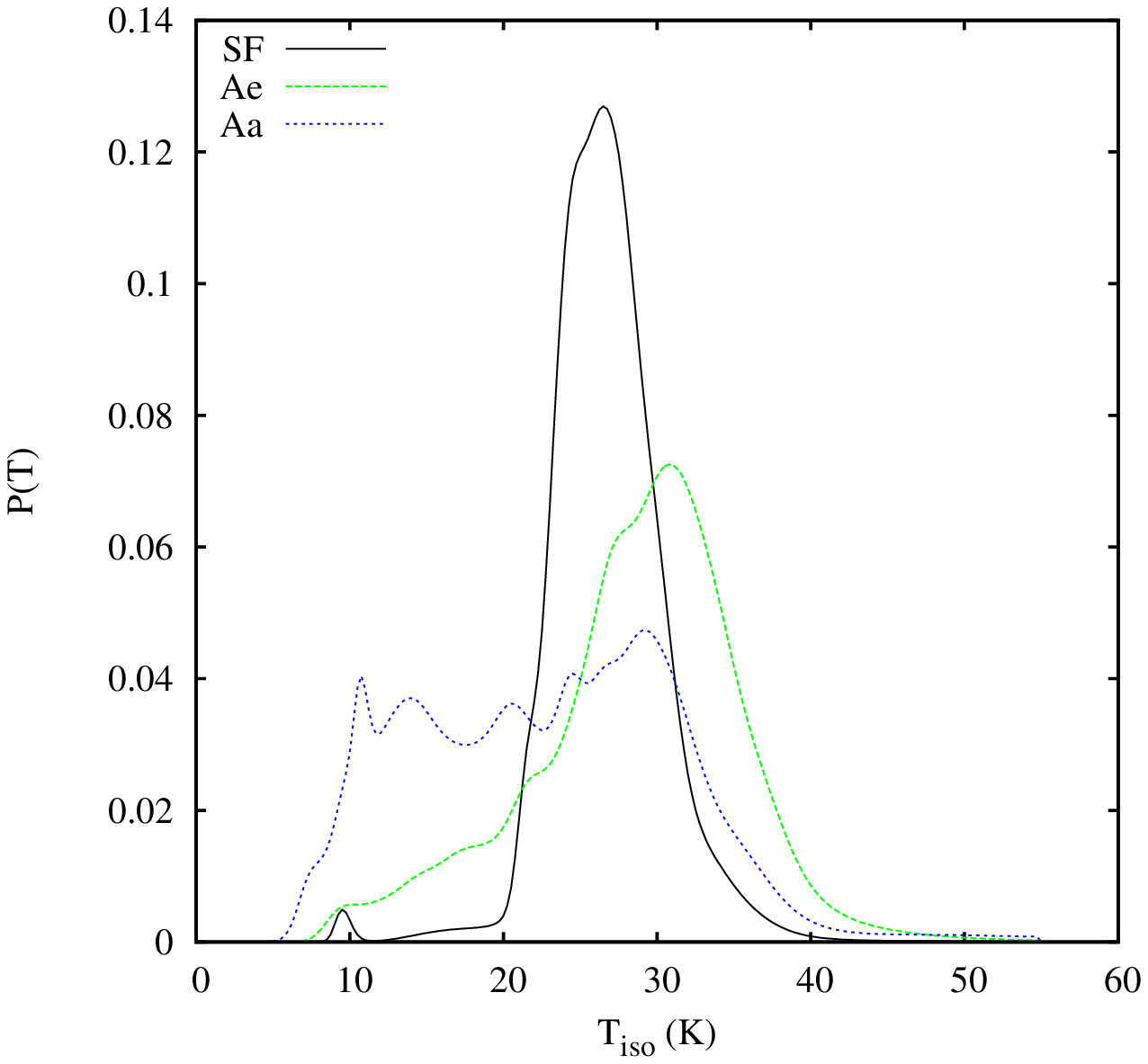}
\caption{Left: a temperature-luminosity plot for the SF,
  Aa and Ae sources with individually determined temperatures.
  Right: stacked, normalized posterior distribution functions (PDFs) for
  temperature fits to the SF, Ae and Aa emission-line classes.}
\label{temps-plot}
\end{figure*}

We began by fitting single-temperature modified black-body models to
all the sources for which this was possible. We selected all objects
which had a $2\sigma$ detection, as defined above, in at least two
{\it Herschel} bands, in order to ensure that there was at least in
principle sufficient signal-to-noise to constrain parameters, and then
used standard Levenberg-Marquardt $\chi^2$ minimization to find the
best-fitting values of temperature and normalization for the modified
blackbody model to the flux densities measured at all five H-ATLAS
bands.

One decision that has to be taken here concerns the emissivity
parameter $\beta$. Earlier work, including H10 and \cite{Dunne+11},
takes this to be 1.5, but work on local galaxies
\citep[e.g.][]{Davies+12} has obtained good fits (using much better
data than available to us) with $\beta=2$. Fitting for $\beta$ across
the whole sample (marginalizing over temperature and normalization for
each object) we find that the best fits are found with $\beta = 1.8$
(the validity of this approach will be discussed by Smith \etal, in prep).
We fixed $\beta$ to this value and then fitted for temperature and
normalization, determining errors by mapping the $\Delta \chi^2 = 2.3$
error ellipse (corresponding to $1\sigma$ for two parameters of
interest). For these final fits, only individual fits with an
acceptable $\chi^2$ value (defined as a reduced $\chi^2 <2$) and a
well-constrained temperature ($\Delta T/T < 2$) are considered in what
follows. This process gives us 385 measured temperatures and
normalizations, including 128 Aa objects (10 per cent of the total),
70 Aes (35 per cent) and 170 SF objects (88 per cent); the `SF cut'
was not applied to the parent sample. Integrated FIR luminosity
($L_{\rm IR}$, integrated between 8 and 1000 microns) and isothermal
dust masses were then calculated from the fitted temperature and
normalization.

Clearly the quantities we measure in this way are expected to be
biased towards the brightest and hottest objects, but it is still
instructive to see how they relate to our emission-line
classifications. Fig.\ \ref{temps-plot} shows the
temperature-luminosity and temperature-dust mass plots for these
objects broken down by emission-line class. We exclude the broad-line
objects, because of concerns noted above about contamination of the
FIR bands by non-thermal emission. We see what appears to be a bimodal
distribution of temperatures, with one set of objects, here seen to be
mostly Aas, having temperatures in the range $10<T<20$ K, while the
other, comprising most of the SF and Ae objects together with a
significant minority of Aas, has $25 < T < 40$ K. The typical error
bar on fitted temperature (not plotted for clarity) is of order 10 per
cent. These temperatures generally seem realistic: the isothermal dust
temperatures measured by \cite{Dunne+11} span the range 10--50 K, and,
unsurprisingly, our temperatures also lie in this range, while
temperatures measured for early-type galaxies in the {\it Herschel}
Reference Survey \citep{SmithM+12} are $\sim 24$ K, which is
comparable to what we see for the Ae sample. What is more surprising
is the cold temperatures found for a number of the Aa objects.
Integrated temperatures of the order of 10 K are perhaps just
plausible for dust in thermal equilibrium with the old stellar
population of elliptical galaxies, if the distribution of dust is
chosen carefully; \cite{Goudfrooij+deJong95} show that temperatures of
tens of K are expected for dust in the centres of elliptical galaxies,
but the effective temperature should be an emission-weighted average
over the dust distribution throughout the galaxy, and the energy
density in photons falls off very rapidly with distance from the
centre of the galaxy, so the integrated temperature depends on dust
distribution, but could be substantially lower than the peak value.
However, the lowest temperatures found by the fits would imply dust
masses of up to $10^{10} M_\odot$, which is probably not realistic.
Inspection of the images for the sources with very low temperatures
suggests that several of them are the result of flux confusion: given
our $2\sigma$ flux cuts, up to 30 of the Aa sources that we have
fitted could be spurious detections, and while we do not expect the
numbers to be this high in practice, confusion seems likely to account
for a number of the sources with the lowest fitted temperatures and
highest dust masses. In addition, synchrotron contamination of the
SPIRE bands, which is observed in nearby LERGs like M87
\citep{Baes+10} could be affecting a few Aas -- several
fall below the dividing line in the plot of Fig.\ \ref{lopezplot},
although not all of these will have the flat integrated radio spectrum
required for non-thermal emission to appear at 500 $\mu$m.

Another way of investigating the temperature differences between the
samples, which does not put so much weight on individual objects, is
to consider the stacked posterior probability distribution over $T$,
marginalizing over normalization, for the fitted objects, and this is
shown in the right-hand panel of Fig.\ \ref{temps-plot}. Here a prior
of $5<T<55$ K is used. We see that this plot reproduces the broad
trends seen in the left-hand panel: the SF objects have a fairly
well-defined peak in temperature at around 26 K, the Aes have a
broader peak at around 30 K, and the Aas span a range between around
10 and 30 K. Even taking into account possible contamination by
confused sources and/or synchrotron emission in a few cases, it does
not seem likely that the Aa and Ae sources have the same intrinsic
temperature distribution. In addition, it is hard to see how this
difference in the PDFs could be explained by, for example, different
$\beta$ values for different populations.

\begin{figure*}
\epsfxsize 8.5cm
\epsfbox{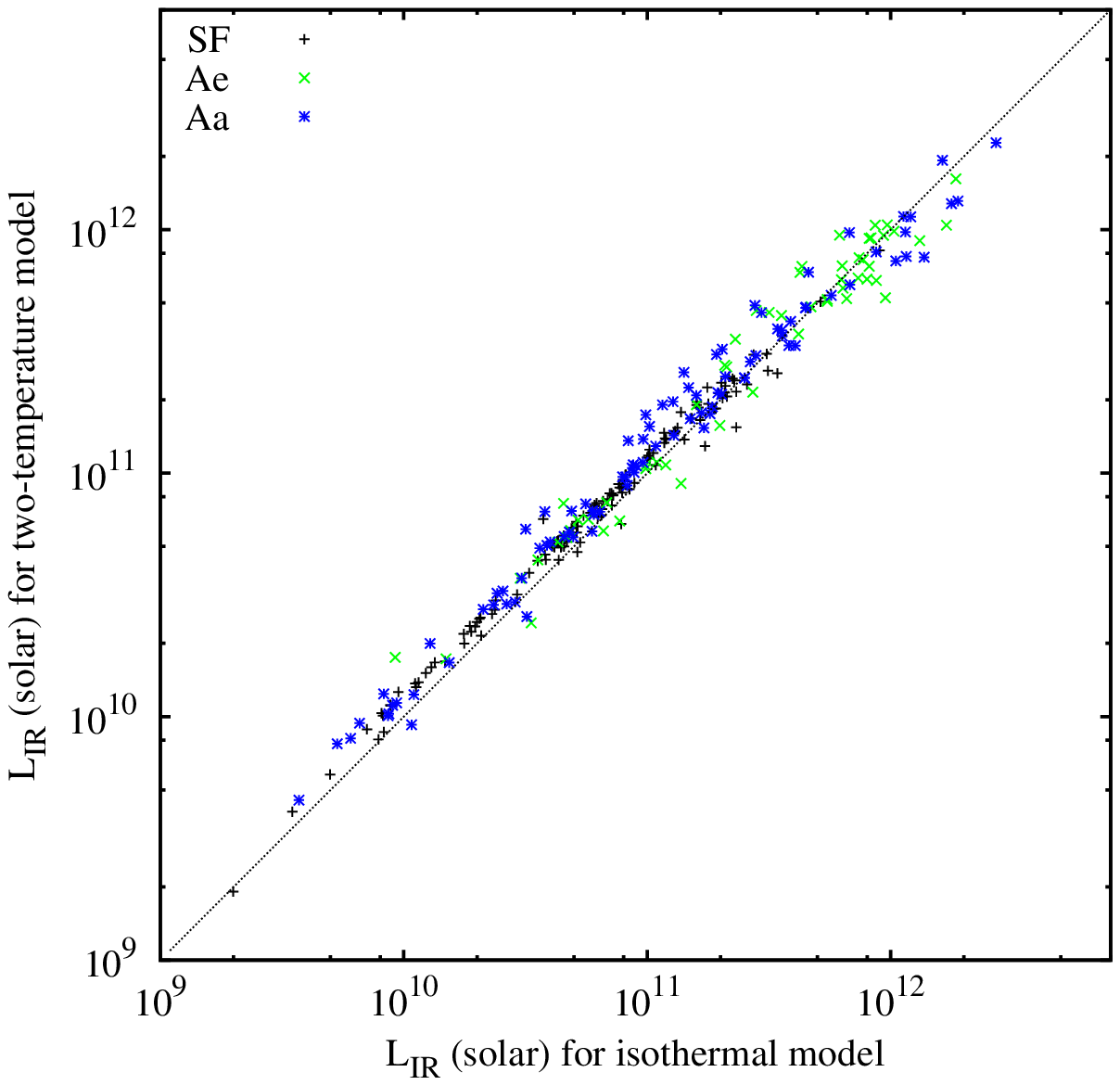}
\epsfxsize 8.5cm
\epsfbox{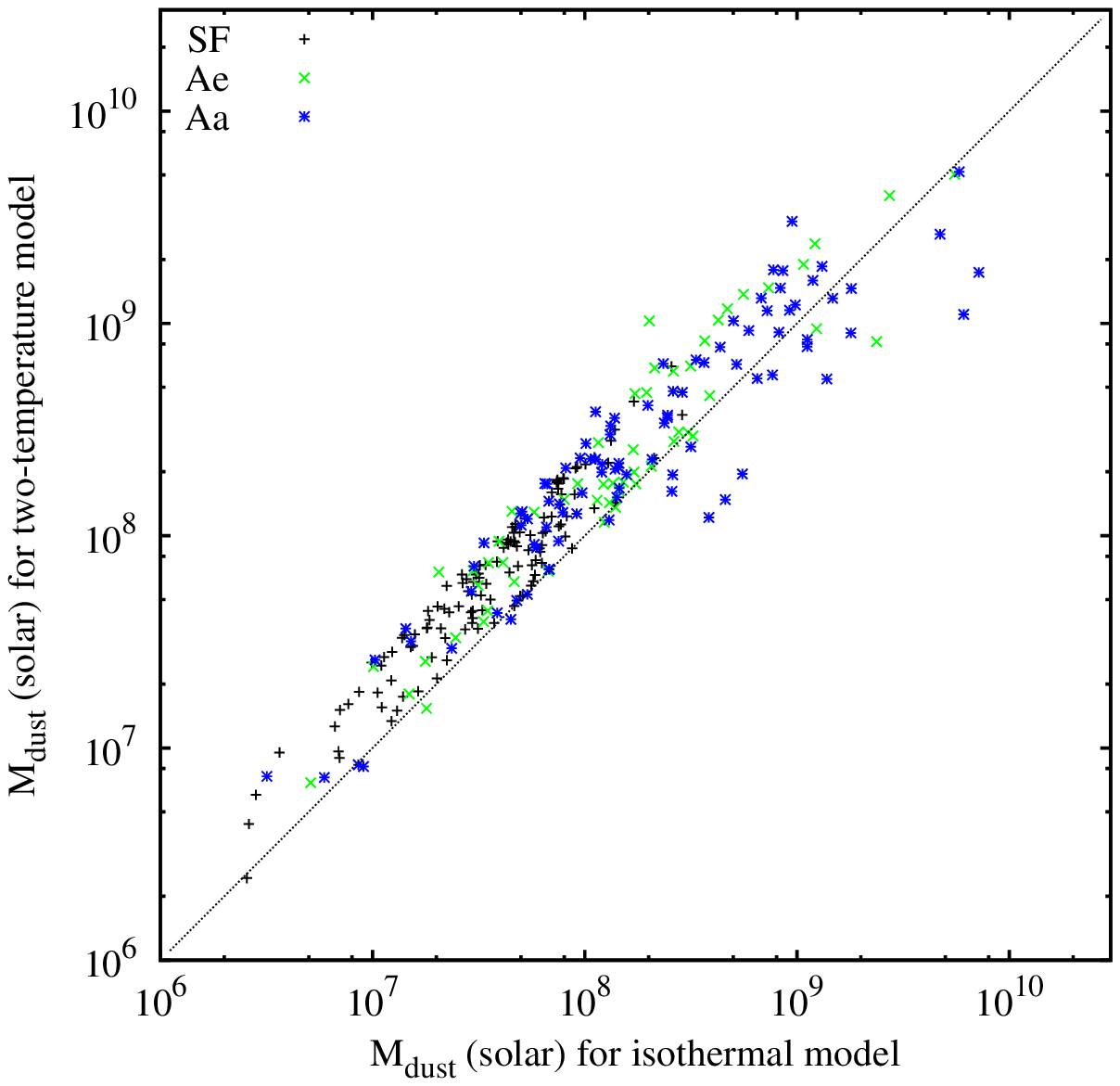}
\caption{The relationships between (left) the integrated IR luminosity
and (right) the dust mass for objects with fits of both a
single-temperature and a two-temperature model.}
\label{twot-plots}
\end{figure*}

It is natural to interpret the wide range of temperatures in the
temperature-luminosity plots in terms of two populations of dust: (1),
a cold dust component which is always present, which is essentially in
thermal equilibrium with the old stellar population ($T \sim 15$ K might
be reasonable for this as an average over the dust properties of an
elliptical galaxy, as noted above) and whose mass scales with the
total galaxy mass, perhaps with some redshift dependence; and (2), a
warmer dust component with $T \sim 30$ K which traces current star
formation and whose mass and luminosity are primarily an indicator of
the star-formation rate \citep[cf.][]{Dunne+11,SmithD+12}. In fact,
such a two-temperature model might help to explain the lack of objects
with $T \sim 20$ K in Fig.\ \ref{temps-plot}: although there should be
objects where the two components contribute roughly equally to the
{\it Herschel} SED, these would tend to be poorly fitted with a
single-temperature model and would be rejected by our fitting
procedure.

The type of broad-band fitting to the FIR through ultraviolet SEDs
carried out by \cite{SmithD+12} is beyond the scope of this paper, but
we investigated a two-temperature model by fitting the same dataset
for the normalizations of two modified-blackbody models with $\beta =
1.8$ and fixed temperatures of 15 K and 30 K (normalizations here are
explicitly constrained to be positive to avoid trading off negative
flux in one component against positive flux in another). These models
in general fit less well and therefore give fewer sources with
acceptable fits, presumably because there are objects with
well-determined temperatures for the warm component that are
significantly different from 30 K, but we do note that they provide
good fits to a population of objects that are rejected by the $\chi^2$
criterion for the single-temperature fits and that generally have
non-zero contributions to the dust luminosity from both hot and cold
components. Moreover, and importantly for what follows, we find a good
correlation between the total luminosities for objects where these can
be obtained using both methods (Fig.\ \ref{twot-plots}), while the
estimated dust masses show some systematic differences. This suggests
that, at least for this sample, the total IR luminosity can be used
without worrying too much about more complex models, while the dust
mass must be interpreted with a little more care. If we interpret the
mass of {\it warm} dust (or, equivalently for this model, its
luminosity) in these fits as tracing star-formation rate, then, using
the results of \citep{SmithD+12}, the typical SF object in our sample
has a star-formation rate of $\sim 10 M_\odot$ yr$^{-1}$, while the
most luminous Ae objects might have star-formation rates more than ten
times higher. More detailed methods for estimating star-formation
rates are discussed in the following subection.

Finally, we note that the weighted mean of the best-fitting single
temperatures of the Aas and Aes, for $\beta = 1.8$, is 20.3 K. This
justifies the assumptions we used for $K$-corrections in the luminosity
stacking of Section \ref{stacking}.

\subsection{$L_{250}$ as a star-formation rate indicator; comparing
  emission-line classes}
\label{SFR}

As we noted in the previous subsection, contributions
to $L_{250}$ are made both by cold dust (driven by the old stellar
population) and warm dust (driven by star formation). It follows that neither
$L_{250}$ nor the integrated $L_{\rm IR}$ are reliable indicators of
star formation rates (SFR) in general. However, they should
both be usable to estimate SFR for an object whose FIR emission is
{\it dominated} by emission from warm dust: these will be the objects
whose best-fitting temperatures are $\sim 25$ K or more. For objects
with a contribution from cold dust, the SFR estimated from $L_{\rm
  250}$ or $L_{\rm IR}$ will be an upper limit.

\begin{figure*}
\hbox{\hskip-1.3cm
\input{sfr-l250}
\hskip-1.2cm
\input{sfr-lir}}
\caption{The relationship between star-formation rate (SFR),
  determined for the SF class from the MPA-JHU emission-line database
  using the methods of \cite{Brinchmann+04}, and the
  two types of FIR luminosity discussed in this paper, $L_{250}$
  (left) and $L_{\rm IR}$ (right). The green lines
show the results of a MCMC regression, as described in the text. Error
bars are not plotted for clarity -- the uncertainty on SFR can be $\pm
0.5$ dex.}
\label{l-sfr}
\end{figure*}
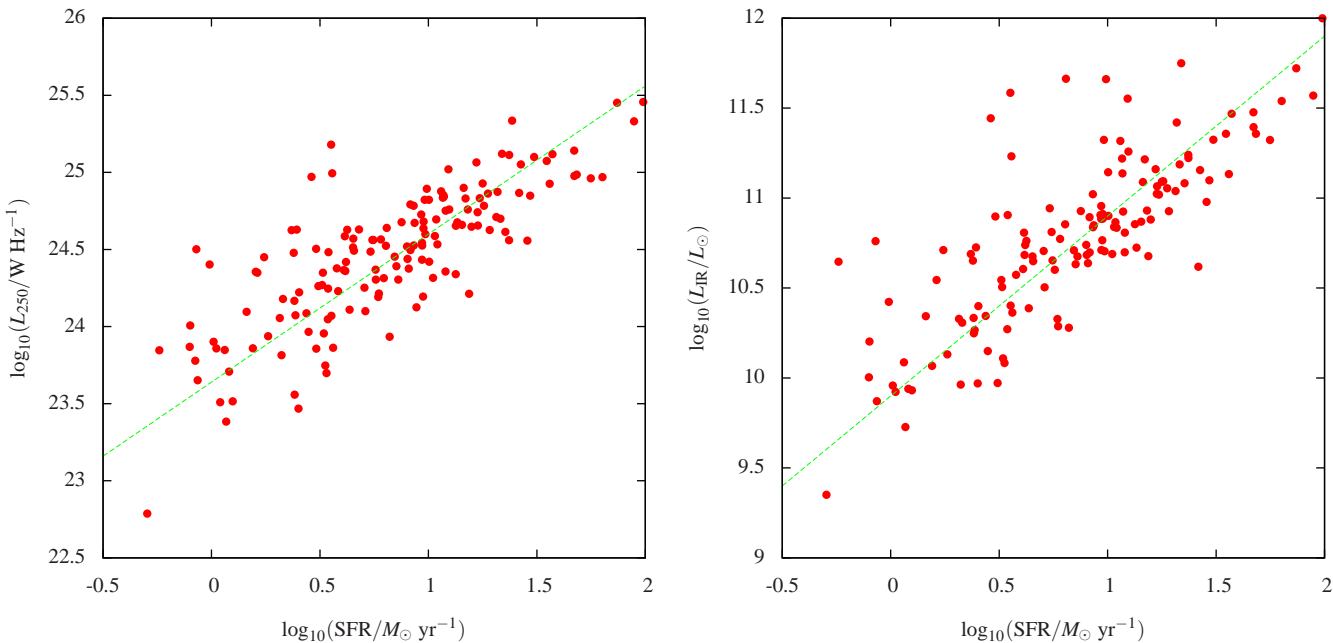

To use the quantity that we have used for stacking, $L_{\rm 250}$, in
this way we need to calibrate the relationship between it and SFR. We
choose to do this by considering the objects classed optically as
`SF', as, where temperature information is available, all of these
have FIR temperatures consistent with being dominated by star
formation (Section \ref{temperatures-ind}). For SF objects with SDSS
spectra estimated star formation rates, derived using the methods of
\cite{Brinchmann+04} (i.e. by model-fitting to the optical emission
lines and stellar continuum), are available in the MPA-JHU database.
Cross-matching our SF objects against this database gives a sample of
158 objects with both SFR and $L_{\rm 250}$ estimates; we use the
median likelihood estimates given in the MPA-JHU database, as being
the most robust, and take half the difference between the 16th and
84th percentiles as an estimate of the error on the SFR. As these
objects are all at low redshifts ($z<0.24$, and median $z = 0.08$) we
need not be concerned about the $K$-correction used to derive $L_{\rm
  250}$. When we plot $L_{\rm 250}$ against SFR derived in this way,
we see a good correlation (Fig.\ \ref{l-sfr}) with a slope that is, by
eye, close to unity. A Markov-Chain Monte Carlo regression, taking the
errors on both SFR and $L_{\rm 250}$ into account and incorporating an
intrinsic dispersion in the manner described by \cite{Hardcastle+09},
gives Bayesian estimates of the slope and intercept of the correlation:
\[
\log_{10}(L_{250}/{\rm W}\ {\rm Hz}^{-1}) = 23.64 + 0.96
\log_{10}({\rm SFR}/M_\odot\ {\rm yr}^{-1})
\]
Although a slope
of unity is not ruled out, we will use this slightly non-linear
relationship in what follows. We emphasise again that it is only valid for
objects whose FIR emission is known to be dominated by warm dust
heated by star formation.

As a sanity check on this approach, we can also estimate the
relationship between SFR and integrated $L_{\rm IR}$ by using our
temperature fits from Section \ref{temperatures-ind}. The vast majority (143)
of SF objects with SFR estimates also have estimates of $T_{\rm IR}$
and thus $L_{\rm IR}$, and this quantity also correlates well with
SFR (Fig.\ \ref{l-sfr}). Regression gives a linear relation
\[
\log_{10}(L_{\rm IR}/L_\odot) = 9.90 + 1.00 \log_{10}({\rm
  SFR}/M_\odot\ {\rm yr}^{-1})
\]
whose normalization is only a factor 1.4 away from the standard
relation given by \cite{Kennicutt98}, derived for starbursts. Thus we
can use our IR-derived SFR with reasonable confidence.

Applying the $L_{250}$/SFR relation to Fig.\ \ref{compar}, we can see
that the most luminous detected radio galaxies, at around $9 \times
10^{25}$ W Hz$^{-1}$, should correspond to star-formation rates around
250 $M_\odot$ yr$^{-1}$, which does not seem unreasonable -- these
would be radio galaxies associated with starbursts -- although it
should be noted that there is a non-negligible uncertainty associated
with the $L_{250}$ values of these luminous, high-$z$ sources because
of the poorly known $K$-correction. Individual powerful, high-$z$ radio galaxies have
  been associated with star formation at levels even higher than this
  \citep{Barthel+12,Seymour+12}. The mean $L_{250}$ of the most
radio-luminous Aes, with $L_{1.4}>10^{25}$ W Hz$^{-1}$, corresponds to
$15M_\odot$ yr$^{-1}$, which is well above the SFRs expected for
normal ellipticals in the local universe. The factor $\sim 4$ between
the stacked $L_{250}$ values for Aes and Aas means that the mean star
formation rate in the latter is {\it at least} a factor 4 below that
in the Aes; `at least' because the temperature measurements suggest
that the emission from some, and perhaps most, Aas is dominated by
cold dust.

\subsection{Stacked dust temperatures, masses and SFR}
\label{temperatures-stacked}

\begin{table*}
\caption{Results of temperature fitting, dust mass and star formation
  rate estimation in
  redshift bins.}
\label{tfit-z-results}
\begin{tabular}{lrrrrrrr}
\hline
Class&Range in&Objects&Best-fit $T$&Reduced&$\log_{10}(L_{250}/{\rm
  W}\ {\rm Hz}^{-1})$&$\log_{10}(M_{\rm iso}/M_\odot)$&$\log_{10}({\rm SFR}/M_\odot\ {\rm yr}^{-1})$\\
&$z$&in bin&(K)&$\chi^2$\\
\hline
Aa& 0.00 -- 0.30 & 398 & $17.0^{+0.9}_{-0.8}$ & 1.65 & $23.9 \pm 0.1$ & $7.6 \pm 0.1$ & $0.2 \pm 0.1$ \\
& 0.30 -- 0.50 & 472 & $17.4^{+0.6}_{-0.6}$ & 1.52 & $24.3 \pm 0.1$ & $8.0 \pm 0.1$ & $0.7 \pm 0.1$ \\
& 0.50 -- 0.90 & 309 & $27.8^{+2.1}_{-3.4}$ & 1.44 & $24.6 \pm 0.1$ & $7.7 \pm 0.1$ & $1.0 \pm 0.1$ \\
Ae& 0.00 -- 0.30 & 62 & $32.8^{+0.7}_{-0.8}$ & 2.11 & $24.3 \pm 0.1$ & $7.3 \pm 0.1$ & $0.7 \pm 0.1$ \\
& 0.30 -- 0.50 & 55 & $30.9^{+0.7}_{-0.8}$ & 1.45 & $24.8 \pm 0.1$ & $7.8 \pm 0.1$ & $1.2 \pm 0.1$ \\
& 0.50 -- 0.90 & 36 & $34.2^{+1.4}_{-1.5}$ & 1.99 & $25.1 \pm 0.1$ & $8.0 \pm 0.1$ & $1.5 \pm 0.1$ \\
\hline
\end{tabular}
\end{table*}

\begin{table*}
\caption{Results of temperature fitting, dust mass and star formation
  rate estimation in
  luminosity bins.}
\label{tfit-l-results}
\begin{tabular}{lrrrrrrr}
\hline
Class&Range in&Objects&Best-fit $T$&Reduced&$\log_{10}(L_{250}/{\rm
  W}\ {\rm Hz}^{-1})$&$\log_{10}(M_{\rm
  iso}/M_\odot)$&$\log_{10}({\rm SFR}/M_\odot\ {\rm yr}^{-1})$\\
&$\log_{10}(L_{1.4})$&in bin&(K)&$\chi^2$\\
\hline
Aa& 22.0 -- 24.0 & 456 & $15.1^{+0.6}_{-0.5}$ & 1.50 & $24.0 \pm 0.1$ & $7.9 \pm 0.1$ & $0.4 \pm 0.1$ \\
& 24.0 -- 25.0 & 589 & $23.8^{+0.8}_{-0.7}$ & 1.65 & $24.5 \pm 0.1$ & $7.7 \pm 0.1$ & $0.9 \pm 0.1$ \\
& 25.0 -- 28.0 & 140 & $15.3^{+0.7}_{-0.6}$ & 1.48 & $24.7 \pm 0.1$ & $8.6 \pm 0.1$ & $1.1 \pm 0.1$ \\
Ae& 22.0 -- 24.0 & 71 & $26.3^{+0.6}_{-0.7}$ & 1.70 & $24.3 \pm 0.1$ & $7.5 \pm 0.1$ & $0.7 \pm 0.1$ \\
& 24.0 -- 25.0 & 57 & $30.9^{+0.5}_{-0.6}$ & 2.08 & $25.0 \pm 0.1$ & $8.0 \pm 0.1$ & $1.5 \pm 0.1$ \\
& 25.0 -- 28.0 & 28 & $27.0^{+1.7}_{-1.8}$ & 1.59 & $24.9 \pm 0.2$ & $8.0 \pm 0.2$ & $1.3 \pm 0.2$ \\
\hline
\end{tabular}
\end{table*}

As noted above, direct estimation of dust temperatures can only be
carried out for the brightest (and possibly hottest) objects, and so
might give misleading results if used in the interpretation of our
stacking analysis of the whole sample. As an alternative, we can
estimate mean temperatures for objects in the sample as follows. We
bin our objects in redshift or radio luminosity as in the previous two
sections. For each redshift/luminosity bin, we determine the {\it
  single} dust temperature that gives the best $\chi^2$ fit to the
observed flux densities of every galaxy in the bin, allowing each galaxy to
have a free normalization (which may be negative) and taking a fixed
$\beta = 1.8$. Errors in this fitted temperature are estimated by
finding the range that gives $\Delta \chi^2 = 1$. We can then use the
best-fitting temperature and normalizations for all the sources to
estimate the 250-$\mu$m luminosity of the bins, determining error bars
by bootstrap as before. The results of this process are tabulated in
Tables \ref{tfit-z-results} and \ref{tfit-l-results}. The $\chi^2$
fitting gives acceptable, though not particularly good results, as we
would expect since, from the analysis of Section
\ref{temperatures-ind}, we know that there is a wide range of
temperatures in each bin. Nevertheless, we can attempt to interpret
the results.

Three points are of interest. Firstly, we note that the luminosities
we estimate are broadly consistent, within the errors, with the
luminosities estimated from the stacking analysis of Section
\ref{stacking}; this gives us confidence that the luminosities from
the earlier analysis are reasonable and that the assumption of a
single temperature for the $K$-correction does not have a big effect
on the inferred monochromatic luminosities. The luminosity difference
between the Aa and Ae spectral classes clearly persists in this
analysis. Secondly, we see that the temperatures are systematically
different for the two emission-line classes: Aes have systematically
higher dust temperatures. Thirdly, we can compute isothermal dust
masses from eq.\ \ref{dustmass} using the best-fitting temperature and
mean luminosity -- these are of course a complicated weighted mean of
the dust masses of all the objects in the bin, but still gives us some
information on the properties of the galaxies. These mean isothermal
dust masses are tabulated for each bin in Tables \ref{tfit-z-results}
and \ref{tfit-l-results}. No very strong difference between the dust
masses for the emission-line classes is seen in these mean masses. It
therefore seems plausible that the clear observed difference in
monochromatic FIR luminosity at 250 $\mu$m between the populations is
driven by a difference in dust temperature rather than by dust mass.

Finally, we can attempt to convert the $L_{\rm 250}$ values from this
fitting into star formation rates, using the results of Section
\ref{SFR}. As already noted, this gives us upper limits if we have
reason to suppose that some of the FIR emission comes from cold dust
unrelated to star formation. The results of this conversion, given in
the final column of Tables \ref{tfit-z-results}
and \ref{tfit-l-results}, must be treated with caution, therefore.
Since the mean fitted temperatures of the Aes are $>30$ K, their SFR
estimates may be a reasonable estimate of the true mean SFR in these
systems; the same is not true of the Aas, and so, again, the safest
interpretation is to say that the mean SFR in the Aes is of order a
few tens of solar masses per year, and is {\it at least}
$\sim 0.5$ dex higher than that in the Aas.

\subsection{Radio source sizes}
\label{length}

\begin{figure*}
\epsfxsize 8.43cm
\epsfbox{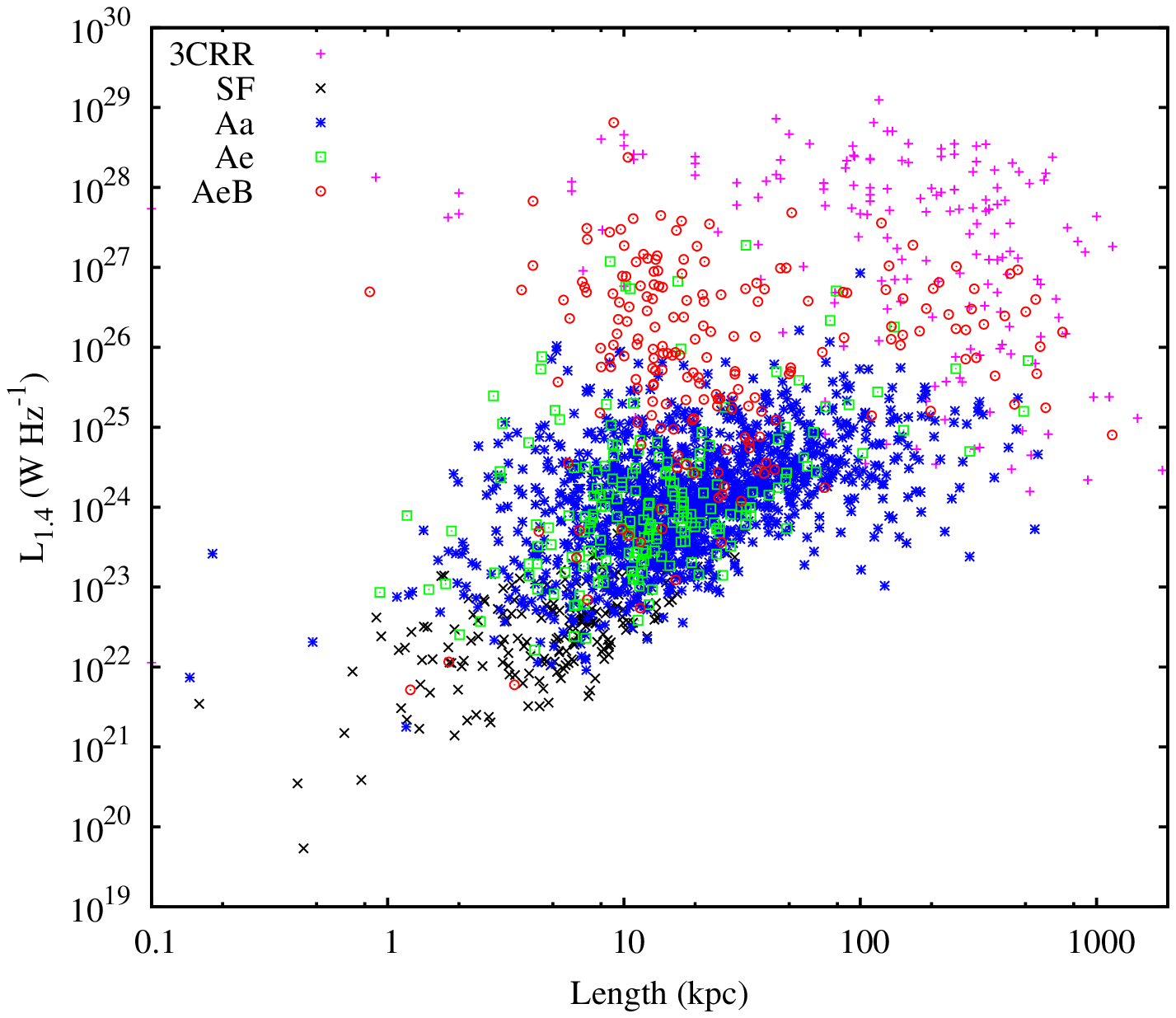}
\epsfxsize 8.57cm
\epsfbox{lenl-j-bins.eps}
\caption{The physical sizes of radio sources in our sample and their
  relationship to radio and FIR luminosity. Left: the
  power/linear-size plot for the sample, broken down by emission-line
  class, and without the SF cut. Colours are as in Fig.\ \ref{llplot};
  for comparison, 3CRR sources are plotted in magenta. Right: the FIR
  luminosity/linear size plot, with stacking in bins at $<40$ and
  $>40$ kpc, after the SF cut. Colours and symbols as in
  Fig.\ \ref{llplot}; upper limits in length are denoted by arrows.}
\label{length-plot}
\end{figure*}

V12 noted a strong relationship between the FIR
luminosity or temperature of the objects they studied and the radio
source size, in the sense that larger objects had systematically lower
$L_{250}$ or $T$. They also showed that this discrepancy was largely
driven by the most massive objects in their sample.

It is clearly interesting to ask how this result relates to the
observed differences between emission-line classes. One of us (JSV)
therefore determined the largest angular size of every object in the
present sample, taking the sizes used by V12 for
objects in common between the two samples and otherwise making
measurements directly from the FIRST images. Where objects were
unresolved in FIRST, an upper limit was assigned, as described by
V12. Scaling by the angular size distance, this gives the distribution
of source physical sizes for the current sample.

An important caveat in this analysis is that the current sample is not
complete. This is illustrated by the left panel of Fig.
\ref{length-plot}, which shows the power/linear-size plot (the `$PD$
diagram') for the current sample together with the equivalent plot for
the complete and well-studied 3CRR sample. For clarity, upper limits
are not marked on this plot, but it should be noted that all sources
with a physical size $\ge 40$ kpc are actually resolved. We see that
all emission-line classes are heavily biased towards smaller physical
sizes with respect to 3CRR: of course, this is not a completely fair
comparison, since the 3CRR objects are the most luminous objects in
the radio sky at any given redshift, and we might expect
lower-luminosity sources to be systematically smaller than the most
luminous ones. However, there are more subtle signs of bias, such as
the fact that there are more large AeB sources than there are Aes:
this arises principally because a broad-line object is more likely to
have a bright radio core and so to be identified with a galaxy or
quasar in our original selection (though there is an additional effect
due to the different redshift distribution of AeBs and Aes). The Aa
sources, which should have the full range of angles to the line of
sight, have a length distribution intermediate between the Aes and
AeBs, as expected. This bias towards compact sources, or sources with
compact cores, is particularly problematic for our sample because of
the selection from FIRST radio images, which resolve out large-scale
emission: the sample of V12 will be closer to being complete.

Having said this, it is still possible to investigate the FIR
properties as a function of length. To do this we apply the SF cut and
then stack the FIR luminosities as in Section \ref{stacking}, binning
by length: we use only two length bins and the division is set at 40
kpc to ensure that all upper limits are in the correct bin. The
results are shown in Fig.\ \ref{length-plot} (right panel). We see
first of all that the Ae/Aa difference persists in this analysis: Ae
sources have higher $L_{250}$ than Aas irrespective of length.
Secondly, we see no evidence for any length dependence of the
$L_{250}$ of the Ae population, although the error bars are large
because the sample is small. Thirdly, we note a marginally significant
difference between the $L_{250}$ values for the small and large Aa
sources, in the sense noted by V12: the null hypothesis that these two
are equal can be rejected at the 95 per cent confidence level. If,
instead of stacking $L_{250}$, we fit temperatures to all the sources
in each bin as described in Section \ref{temperatures-stacked}, we
find {\it no} significant difference in luminosities as a function of
length for either emission-line class, but there are significant
differences in best-fitting temperature ($29.4 \pm 0.5$ vs $22 \pm 2$
K for Ae; $21.8 \pm 0.5$ vs $12.2 \pm 0.4$ K for Aa), in both cases in
the sense that the larger objects have lower best-fitting
temperatures. This is again broadly consistent with the results of
V12.

\section{Discussion}
\label{discussion}

The results of the previous section show very clearly that there is a
difference between the average far-infrared properties of radio
galaxies whose spectra show strong emission lines and those of radio
galaxies that do not. How can we interpret this?

It is first of all important to consider the issue of possible AGN
contamination, discussed in Section \ref{intro}, in more detail. There
are two possible sources for this: (1) emission from the warm dusty
torus, which is expected to be seen predominantly in HERGs, and (2)
synchrotron emission from the jets and lobes, which may appear in all
objects. The first of these is particularly important, as
torus-related emission in the FIR bands might give rise to a HERG/LERG
difference, but we are confident that it is not a significant effect
in our sample, for several reasons. Firstly, when the required mid-IR
data are available, which is not the case for our sample at present,
decompositions of the SEDs of radio-loud and radio-quiet AGN tend to
show that observer-frame {\it Herschel} SPIRE bands are dominated by
cool dust rather than by the torus component, even for powerful AGN
with luminous tori \citep[e.g][]{Barthel+12,DelMoro+12}. Secondly, we
know, as pointed out by H10, that the mid-IR torus luminosities even
for the most powerful HERGs in our sample, if they follow the
correlation between radio power and mid-IR luminosity established by
\cite{Hardcastle+09}, should be about two orders of magnitude less
than the total FIR luminosities estimated e.g. in
Fig.\ \ref{temps-plot}, implying that even if the torus SEDs were
strikingly different from those of known radio-loud AGN, it would
still be energetically impossible for them to affect the observed FIR
emission significantly. The second possible source of contamination,
synchrotron emission, we believe to affect mostly the broad-line
objects, as discussed in Section \ref{blrg}, together with at most a
very few of the Aas; there is certainly no reason to expect that it
would give rise to the observed Aa/Ae difference, since the radio
fluxes and spectra of these two classes are very similar. We therefore
consider it safe to discuss our observations in terms of a difference
  in the properties of cool dust in the two populations.

Our result cannot be significantly affected by contamination by pure
star-forming objects whose radio emission is bright enough to cause
them to be misidentified as radio galaxies. While our spectroscopic
classification alone does not identify all objects whose radio
emission is dominated by star formation (Fig.\ \ref{llplot}) the
combination of optical spectroscopy and the `SF cut' that we impose on
the radio-FIR luminosity plane, where a clear star-forming sequence is
visible, should remove all such objects. Moreover, the highest radio
luminosities in our sample are well above even the $\sim 10^{25}$ W
Hz$^{-1}$ expected from a starburst of a few thousand $M_\odot$
yr$^{-1}$, and we see a clear difference between the different
emission-line classes in this luminosity range.

Along similar lines, we do not believe that the relationship between
emission-line class and FIR emission can be a result of optical
emission-line activity due to the star formation process itself. Our
emission-line classification uses [O{\sc iii}], which, at least at
high luminosities, is widely used as an AGN indicator, although it can
be produced by hot, young stars. We do not have a direct estimate of
the [O{\sc iii}] luminosities of our sample objects in the version of
the GANDALF-derived database we use, but we have derived a rough
indicator of luminosity from the measured equivalent widths and the
$K$-corrected absolute magnitude in the SDSS $g$ band. Calibrating
this indicator using the MPA-JHU emission-line measurements, for which
both equivalent width and [O{\sc iii}] flux are tabulated, we see that
almost all the Ae objects would be expected to have $L([\rm{O\hskip
    0.02cm\scriptscriptstyle{III}}])>10^{40}$ erg s$^{-1}$, and are
thus in the range classified by e.g. \cite{Kauffmann+03} as `strong
AGN'. Further work in this area will require measurements of the
[O{\sc iii}] fluxes, and ideally those of other lines, for a large
radio-galaxy sample, but we are confident that we are assessing
genuine AGN activity in the vast majority of cases.

We can therefore move on to interpreting the relationship as being one
between FIR properties of the host galaxy and the AGN-related
emission-line properties of radio galaxies that we discussed in
Section \ref{intro}, with the Aa population corresponding to LERGs and
the Ae population to HERGs. It then appears that HERGs, on average,
have significantly higher $L_{250}$ than LERGs (Section
\ref{stacking}); moreover, HERGs appear to have higher $L_{250}$ than
normal galaxies of comparable absolute magnitude at all redshifts
(Section \ref{normal}). In a simple isothermal model, higher
luminosity can arise either because of higher masses of dust or higher
temperatures; what we see from the analysis of Sections
\ref{temperatures-ind} and \ref{temperatures-stacked} is that it is
plausible that the dust masses of the different systems are similar,
but that the mean isothermal temperatures of the HERGs are higher. In
resolved local galaxies, it has been shown that low-temperature dust
emission ($T \sim 15$ K) is driven by the old stellar population,
while significantly hotter temperatures are seen from star-forming
regions \citep[e.g.][]{Bendo+10,Boquien+11}. By far the most obvious
interpretation of our result is therefore that the star-formation
rates are significantly higher in the HERG subsamples than in the
LERGs, giving rise to a significant component of emission from hot
dust which raises the isothermal temperature as seen in the analysis
of \cite{Dunne+11}. If so, this is strong confirmation, using a much
larger sample, of the picture that emerges from the earlier work
discussed in Section \ref{intro}
\citep[e.g.][]{Baldi+Capetti08,Herbert+10,RamosAlmeida+11b,RamosAlmeida+12}.
By calibrating $L_{\rm 250}$ as a star formation indicator using SFRs
derived from local radio-loud star-forming galaxies (Section
\ref{SFR}) we have been able to quantify this, showing that the mean
SFR in the most luminous/high-$z$ Aes is probably at the level of
around $30 M_\odot$ yr$^{-1}$, and is {\it at least} $\sim 0.5$ dex
higher than that in the Aas at all redshifts and radio luminosities.

What does this tell us about the association between star formation
and AGN activity in radio galaxies? The first point to note is that
the association between a HERG classification and increased FIR
luminosity (and therefore star-formation rate) is statistical only. It
can be seen from Fig.\ \ref{compar} that there are individual LERGs
with high FIR luminosities, while at the same radio luminosity we see
HERGs with FIR luminosities 0.5-1 dex lower. Similarly, the
temperature analysis of Section \ref{temperatures-ind} shows that
there are LERGs with best-fitting temperatures comparable (within the
large errors) to those of the warm dust in known star-forming
galaxies. Nothing appears to require HERGs to be associated with high
star-formation rates or LERGs to be associated with completely
quiescent galaxies, consistent with the conclusions of
\cite{Tadhunter+11}. This suggests that the mechanism of the
association is not the simplest possible one, in which some single
event, such as a merger, always triggers both HERG activity and star
formation. If this were the case, we would not see individual LERGs
with high star-formation rates (setting aside the possibility, which
we regard as remote, that these sources are all misidentified HERGs).

Another piece of evidence supporting this picture comes from the lobe
length analysis of Section \ref{length}. If HERGs were associated with
AGN triggering following a merger, we might expect them to show a very
strong relationship between FIR properties and lobe physical size, since star
formation would be expected to peak on average at early times in the radio
source's lifetime. This was a possible interpretation of the results
of V12, who showed that larger sources in general have lower FIR
temperatures and luminosities. However, our analysis shows that this
result is {\it not} driven by the HERG (Ae) population, and in fact for
our sample is more obvious for the LERGs (which, however, have
considerably better statistics).

We would therefore argue that we are not seeing a simple triggering
relationship, but rather that the difference between HERGs and LERGs
is that HERGs tend to inhabit environments in which star formation is
favoured relative to the general galaxy population, while by contrast
star formation is disfavoured in the environments of LERGs. The FIR
differences as a function of source length would then be explained by
some other process, such as jet-induced star formation when the bow
shock of the source is within the host galaxy, which can in principle
take place in both emission-line classes (although we note there is
not yet any direct evidence for this process affecting emission seen
in the FIR band). Such a model is consistent both with all the
observations to date
\citep[e.g.][]{Baldi+Capetti08,Herbert+10,Best+Heckman12,Janssen+12}
and with the explanation of the HERG/LERG dichotomy in terms of
accretion mode discussed in Section \ref{intro}. It will be of great
interest to see whether this result is confirmed by the larger samples
that will be made available by the full H-ATLAS dataset, whether it
can be extended to higher redshifts using deeper spectroscopic or
photometric surveys, and whether the same results are obtained when
the HERG/LERG classification is made using data at other wavebands
(e.g. X-ray or mid-IR).

\section{Summary and conclusions}

The key results from the analysis and discussion above can be
summarized as follows:
\begin{itemize}
\item We have used individual measurements and stacking analyses to
  determine the FIR properties (mean luminosities and temperatures) of
  a large sample of radio-selected sources with spectroscopic redshifts
  and HERG/LERG classifications from optical spectroscopy. Sources near
  the known FIR-radio correlation are excluded from our analysis; the
  vast majority of the objects we study should be bona fide radio galaxies.
\item We find a clear difference between the FIR properties of the two
  populations in the sense that the rest-frame 250-$\mu$m luminosities
  are systematically higher in the HERGs than in the LERGs; the host
  galaxies of LERGs in fact occupy galaxies with lower FIR luminosities
  than normal galaxies matched in absolute magnitude, while HERGs tend
  to have higher FIR luminosities. This difference is apparent at all
  redshifts and all radio luminosities sampled by our targets.
\item A comparison of the temperatures and dust masses of HERGs and
  LERGs, stacked in coarse bins, suggests that the dust masses are
  reasonably comparable for the two samples but that the temperatures
  in the HERGs are systematically higher. This provides strong evidence
  that the higher FIR luminosities we are seeing imply, on average,
  higher star formation rates (which are required to raise the mean
  temperature of the dust) rather than just higher dust masses. The
  low mean temperatures seen for LERGs are consistent with what would
  be expected for quiescent dust which is in thermal equilibrium with
  the photon field of the old stellar population of the host galaxy,
  although the fact that these objects are detected at all implies
  that large masses of dust are present.
\item Quantifying the SFR by calibrating $L_{250}$ as a star-formation
  indicator in the `SF' sources known to be dominated by hot dust, we
  find that the mean SFR in the radio-luminous Aes is $\sim 30M_\odot$
  yr$^{-1}$, and is {\it at least} $\sim 0.5$ dex higher than that in
  the Aas at all luminosities and redshifts.
\item Consistent with the results of V12, we find that both
  emission-line classes in our sample show some evidence for a
  dependence of FIR properties on radio source size.
\item We argue that there is certainly not a simple triggering
  relation, and not even a one-to-one association, between enhanced
  star formation and a particular AGN type \citep[a conclusion
    consistent with detailed studies of starburst radio galaxies such
    as that of][]{Tadhunter+11}. However, the statistical trend for
  HERGs to have higher star formation rates is consistent both with
  what is known from other wavebands
  \citep[e.g.][]{Best+Heckman12,Janssen+12} and with the general class
  of models \citep{Hardcastle+07-2} in which HERG activity takes place
  in lower-mass galaxies where the black hole is able to accrete
  significant quantities of cold gas.
\end{itemize}

As more H-ATLAS data and supporting optical imaging and spectroscopy
become available we expect to extend this work to much larger samples,
allowing more detailed binning and temperature analysis, to
investigate different methods of carrying out the LERG/HERG
classification, and to consider radio galaxies at higher redshifts in
order to search for evidence of cosmological evolution of the
star-formation properties of the radio-loud AGN population.

\section*{Acknowledgements}

The Herschel-ATLAS is a project with Herschel, which is an ESA space
observatory with science instruments provided by European-led
Principal Investigator consortia and with important participation from
NASA. The H-ATLAS website is http://www.h-atlas.org/.

GAMA is a joint
European-Australasian project, based around a spectroscopic campaign
using the AAOmega instrument, and is funded by the STFC, the ARC, and
the AAO. The GAMA input catalogue is based on data taken from the
Sloan Digital Sky Survey and the UKIRT Infrared Deep Sky Survey.
Complementary imaging of the GAMA regions is being obtained by a
number of independent survey programs including GALEX MIS, VST KIDS,
VISTA VIKING, WISE, Herschel-ATLAS, GMRT and ASKAP providing UV to
radio coverage. GAMA is funded by the STFC (UK), the ARC (Australia),
the AAO, and the participating institutions. The GAMA website is
http://www.gama-survey.org/. Funding for the SDSS and SDSS-II has been
provided by the Alfred P. Sloan Foundation, the Participating
Institutions, the National Science Foundation, the U.S. Department of
Energy, the National Aeronautics and Space Administration, the
Japanese Monbukagakusho, the Max Planck Society, and the Higher
Education Funding Council for England. The SDSS website is
http://www.sdss.org/.

The National Radio Astronomy Observatory (NRAO)
is a facility of the National Science Foundation operated under
cooperative agreement by Associated Universities, Inc.

We thank an anonymous referee for helpful comments on the paper.

\bibliographystyle{mn2e}
\renewcommand{\refname}{REFERENCES}
\setlength{\bibhang}{2.0em}
\setlength\labelwidth{0.0em}
\bibliography{mjh,cards}

\end{document}

%% file: sfr-l250.tex
\begingroup
  \makeatletter
  \providecommand\color[2][]{%
    \GenericError{(gnuplot) \space\space\space\@spaces}{%
      Package color not loaded in conjunction with
      terminal option `colourtext'%
    }{See the gnuplot documentation for explanation.%
    }{Either use 'blacktext' in gnuplot or load the package
      color.sty in LaTeX.}%
    \renewcommand\color[2][]{}%
  }%
  \providecommand\includegraphics[2][]{%
    \GenericError{(gnuplot) \space\space\space\@spaces}{%
      Package graphicx or graphics not loaded%
    }{See the gnuplot documentation for explanation.%
    }{The gnuplot epslatex terminal needs graphicx.sty or graphics.sty.}%
    \renewcommand\includegraphics[2][]{}%
  }%
  \providecommand\rotatebox[2]{#2}%
  \@ifundefined{ifGPcolor}{%
    \newif\ifGPcolor
    \GPcolortrue
  }{}%
  \@ifundefined{ifGPblacktext}{%
    \newif\ifGPblacktext
    \GPblacktexttrue
  }{}%
  \let\gplgaddtomacro\g@addto@macro
  \gdef\gplbacktext{}%
  \gdef\gplfronttext{}%
  \makeatother
  \ifGPblacktext
    \def\colorrgb#1{}%
    \def\colorgray#1{}%
  \else
    \ifGPcolor
      \def\colorrgb#1{\color[rgb]{#1}}%
      \def\colorgray#1{\color[gray]{#1}}%
      \expandafter\def\csname LTw\endcsname{\color{white}}%
      \expandafter\def\csname LTb\endcsname{\color{black}}%
      \expandafter\def\csname LTa\endcsname{\color{black}}%
      \expandafter\def\csname LT0\endcsname{\color[rgb]{1,0,0}}%
      \expandafter\def\csname LT1\endcsname{\color[rgb]{0,1,0}}%
      \expandafter\def\csname LT2\endcsname{\color[rgb]{0,0,1}}%
      \expandafter\def\csname LT3\endcsname{\color[rgb]{1,0,1}}%
      \expandafter\def\csname LT4\endcsname{\color[rgb]{0,1,1}}%
      \expandafter\def\csname LT5\endcsname{\color[rgb]{1,1,0}}%
      \expandafter\def\csname LT6\endcsname{\color[rgb]{0,0,0}}%
      \expandafter\def\csname LT7\endcsname{\color[rgb]{1,0.3,0}}%
      \expandafter\def\csname LT8\endcsname{\color[rgb]{0.5,0.5,0.5}}%
    \else
      \def\colorrgb#1{\color{black}}%
      \def\colorgray#1{\color[gray]{#1}}%
      \expandafter\def\csname LTw\endcsname{\color{white}}%
      \expandafter\def\csname LTb\endcsname{\color{black}}%
      \expandafter\def\csname LTa\endcsname{\color{black}}%
      \expandafter\def\csname LT0\endcsname{\color{black}}%
      \expandafter\def\csname LT1\endcsname{\color{black}}%
      \expandafter\def\csname LT2\endcsname{\color{black}}%
      \expandafter\def\csname LT3\endcsname{\color{black}}%
      \expandafter\def\csname LT4\endcsname{\color{black}}%
      \expandafter\def\csname LT5\endcsname{\color{black}}%
      \expandafter\def\csname LT6\endcsname{\color{black}}%
      \expandafter\def\csname LT7\endcsname{\color{black}}%
      \expandafter\def\csname LT8\endcsname{\color{black}}%
    \fi
  \fi
  \setlength{\unitlength}{0.0500bp}%
  \begin{picture}(5760.00,5040.00)%
    \gplgaddtomacro\gplbacktext{%
      \csname LTb\endcsname%
      \put(1210,704){\makebox(0,0)[r]{\strut{} 22.5}}%
      \put(1210,1286){\makebox(0,0)[r]{\strut{} 23}}%
      \put(1210,1867){\makebox(0,0)[r]{\strut{} 23.5}}%
      \put(1210,2449){\makebox(0,0)[r]{\strut{} 24}}%
      \put(1210,3030){\makebox(0,0)[r]{\strut{} 24.5}}%
      \put(1210,3612){\makebox(0,0)[r]{\strut{} 25}}%
      \put(1210,4193){\makebox(0,0)[r]{\strut{} 25.5}}%
      \put(1210,4775){\makebox(0,0)[r]{\strut{} 26}}%
      \put(1342,484){\makebox(0,0){\strut{}-0.5}}%
      \put(2159,484){\makebox(0,0){\strut{} 0}}%
      \put(2977,484){\makebox(0,0){\strut{} 0.5}}%
      \put(3794,484){\makebox(0,0){\strut{} 1}}%
      \put(4612,484){\makebox(0,0){\strut{} 1.5}}%
      \put(5429,484){\makebox(0,0){\strut{} 2}}%
      \put(708,2739){\rotatebox{-270}{\makebox(0,0){\strut{}$\log_{10}(L_{250}/\mathrm{W}\ \mathrm{Hz}^{-1})$}}}%
      \put(3385,154){\makebox(0,0){\strut{}$\log_{10}(\mathrm{SFR}/M_\odot\ \mathrm{yr}^{-1})$}}%
    }%
    \gplgaddtomacro\gplfronttext{%
    }%
    \gplbacktext
    \put(0,0){\includegraphics{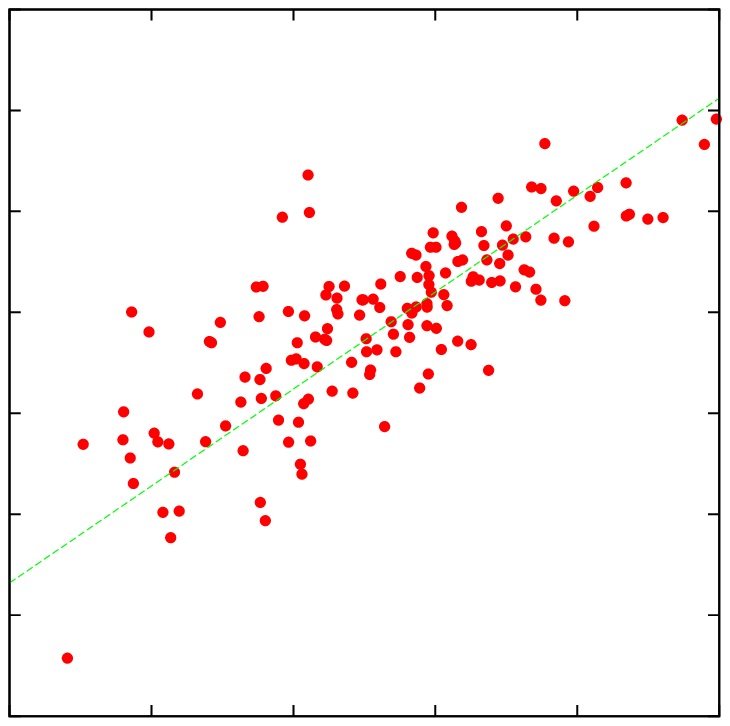}}%
    \gplfronttext
  \end{picture}%
\endgroup

%% file: sfr-lir.tex
\begingroup
  \makeatletter
  \providecommand\color[2][]{%
    \GenericError{(gnuplot) \space\space\space\@spaces}{%
      Package color not loaded in conjunction with
      terminal option `colourtext'%
    }{See the gnuplot documentation for explanation.%
    }{Either use 'blacktext' in gnuplot or load the package
      color.sty in LaTeX.}%
    \renewcommand\color[2][]{}%
  }%
  \providecommand\includegraphics[2][]{%
    \GenericError{(gnuplot) \space\space\space\@spaces}{%
      Package graphicx or graphics not loaded%
    }{See the gnuplot documentation for explanation.%
    }{The gnuplot epslatex terminal needs graphicx.sty or graphics.sty.}%
    \renewcommand\includegraphics[2][]{}%
  }%
  \providecommand\rotatebox[2]{#2}%
  \@ifundefined{ifGPcolor}{%
    \newif\ifGPcolor
    \GPcolortrue
  }{}%
  \@ifundefined{ifGPblacktext}{%
    \newif\ifGPblacktext
    \GPblacktexttrue
  }{}%
  \let\gplgaddtomacro\g@addto@macro
  \gdef\gplbacktext{}%
  \gdef\gplfronttext{}%
  \makeatother
  \ifGPblacktext
    \def\colorrgb#1{}%
    \def\colorgray#1{}%
  \else
    \ifGPcolor
      \def\colorrgb#1{\color[rgb]{#1}}%
      \def\colorgray#1{\color[gray]{#1}}%
      \expandafter\def\csname LTw\endcsname{\color{white}}%
      \expandafter\def\csname LTb\endcsname{\color{black}}%
      \expandafter\def\csname LTa\endcsname{\color{black}}%
      \expandafter\def\csname LT0\endcsname{\color[rgb]{1,0,0}}%
      \expandafter\def\csname LT1\endcsname{\color[rgb]{0,1,0}}%
      \expandafter\def\csname LT2\endcsname{\color[rgb]{0,0,1}}%
      \expandafter\def\csname LT3\endcsname{\color[rgb]{1,0,1}}%
      \expandafter\def\csname LT4\endcsname{\color[rgb]{0,1,1}}%
      \expandafter\def\csname LT5\endcsname{\color[rgb]{1,1,0}}%
      \expandafter\def\csname LT6\endcsname{\color[rgb]{0,0,0}}%
      \expandafter\def\csname LT7\endcsname{\color[rgb]{1,0.3,0}}%
      \expandafter\def\csname LT8\endcsname{\color[rgb]{0.5,0.5,0.5}}%
    \else
      \def\colorrgb#1{\color{black}}%
      \def\colorgray#1{\color[gray]{#1}}%
      \expandafter\def\csname LTw\endcsname{\color{white}}%
      \expandafter\def\csname LTb\endcsname{\color{black}}%
      \expandafter\def\csname LTa\endcsname{\color{black}}%
      \expandafter\def\csname LT0\endcsname{\color{black}}%
      \expandafter\def\csname LT1\endcsname{\color{black}}%
      \expandafter\def\csname LT2\endcsname{\color{black}}%
      \expandafter\def\csname LT3\endcsname{\color{black}}%
      \expandafter\def\csname LT4\endcsname{\color{black}}%
      \expandafter\def\csname LT5\endcsname{\color{black}}%
      \expandafter\def\csname LT6\endcsname{\color{black}}%
      \expandafter\def\csname LT7\endcsname{\color{black}}%
      \expandafter\def\csname LT8\endcsname{\color{black}}%
    \fi
  \fi
  \setlength{\unitlength}{0.0500bp}%
  \begin{picture}(5760.00,5040.00)%
    \gplgaddtomacro\gplbacktext{%
      \csname LTb\endcsname%
      \put(1210,704){\makebox(0,0)[r]{\strut{} 9}}%
      \put(1210,1383){\makebox(0,0)[r]{\strut{} 9.5}}%
      \put(1210,2061){\makebox(0,0)[r]{\strut{} 10}}%
      \put(1210,2740){\makebox(0,0)[r]{\strut{} 10.5}}%
      \put(1210,3418){\makebox(0,0)[r]{\strut{} 11}}%
      \put(1210,4097){\makebox(0,0)[r]{\strut{} 11.5}}%
      \put(1210,4775){\makebox(0,0)[r]{\strut{} 12}}%
      \put(1342,484){\makebox(0,0){\strut{}-0.5}}%
      \put(2159,484){\makebox(0,0){\strut{} 0}}%
      \put(2977,484){\makebox(0,0){\strut{} 0.5}}%
      \put(3794,484){\makebox(0,0){\strut{} 1}}%
      \put(4612,484){\makebox(0,0){\strut{} 1.5}}%
      \put(5429,484){\makebox(0,0){\strut{} 2}}%
      \put(708,2739){\rotatebox{-270}{\makebox(0,0){\strut{}$\log_{10}(L_\mathrm{IR}/L_\odot)$}}}%
      \put(3385,154){\makebox(0,0){\strut{}$\log_{10}(\mathrm{SFR}/M_\odot\ \mathrm{yr}^{-1})$}}%
    }%
    \gplgaddtomacro\gplfronttext{%
    }%
    \gplbacktext
    \put(0,0){\includegraphics{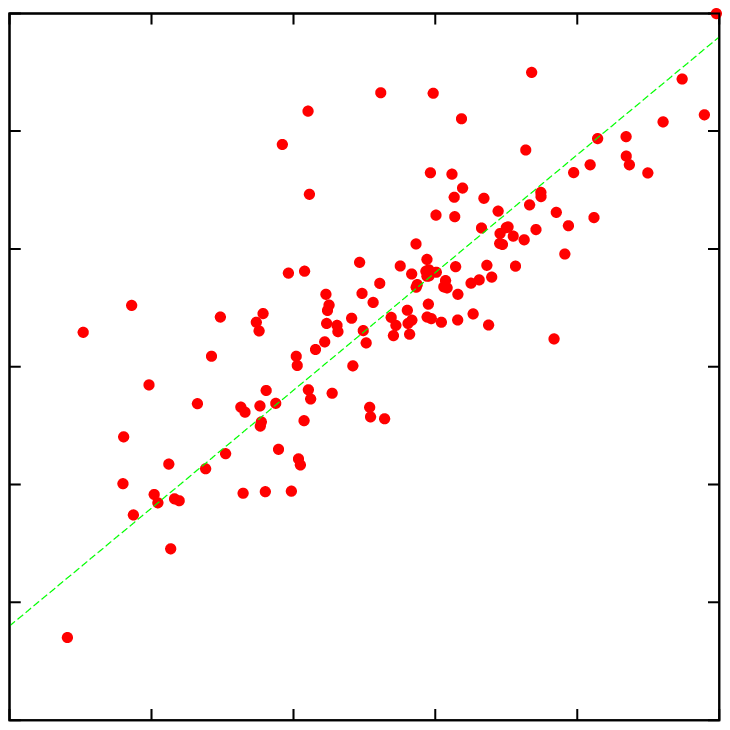}}%
    \gplfronttext
  \end{picture}%
\endgroup